\pdfoutput=1
\documentclass{article}

\usepackage{graphicx}
\usepackage{amsmath}
\usepackage{amssymb}
\usepackage{dcolumn}
\usepackage{bm}
\usepackage{slashed}
\usepackage{authblk}
\usepackage{hyperref}

\newcommand{\be}{\begin{equation}}
\newcommand{\ee}{\end{equation}}
\newcommand{\ba}{\begin{eqnarray}}
\newcommand{\ea}{\end{eqnarray}}
\newcommand{\no}{\nonumber \\}
\newcommand{\gsim}{\mathrel{\hbox{\rlap{\lower.55ex \hbox {$\sim$}}
                   \kern-.3em \raise.4ex \hbox{$>$}}}}
\newcommand{\lsim}{\mathrel{\hbox{\rlap{\lower.55ex \hbox {$\sim$}}
                   \kern-.3em \raise.4ex \hbox{$<$}}}}

\def\roughly#1{\mathrel{\raise.3ex\hbox{$#1$\kern-.75em%
\lower1ex\hbox{$\sim$}}}}
\def\lsim{\roughly<}
\def\gsim{\roughly>}

\def\hb{\hbar}
\def\({\left(}
\def\){\right)}
\def\[{\left[}
\def\]{\right]}
\def\<{\langle}
\def\>{\rangle}
\def\bh{{\bf h}}
\def\bppp{{\bf p}_\perp}
\def\bkpp{{\bf k}_\perp}

\def\k{{\kappa}}
\def\l{{\lambda}}
\def\L{{\Lambda}}
\def\d{{\delta}}
\def\D{{\Delta}}

\def\O{{\Omega}}
\def\e{{\epsilon}}

\def\a{{\alpha}}
\def\b{{\beta}}
\def\c{{\chi}}

\def\g{{\gamma}}
\def\G{{\Gamma}}
\def\h{\eta}
\def\p{{\pi}}
\def\P{{\Pi}}
\def\m{{\mu}}
\def\n{{\nu}}
\def\r{{\rho}}
\def\s{{\sigma}}
\def\S{{\Sigma}}
\def\t{{\tau}}
\def\th{{\theta}}

\def\ps{{\psi}}

\def\x{{\xi}}
\def\P{{\Pi}}

\newcommand{\pd}{{\partial}}
\newcommand{\dg}{\dagger}
\newcommand{\pr}{\parallel}

\newcommand{\pb}{\text{PB}}
\newcommand{\tr}{\text{tr}}
\newcommand{\Tr}{\text{Tr}}
\newcommand{\pp}{\perp}
\newcommand{\he}{\hat{e}}
\newcommand{\ppr}{p_{\parallel}}
\newcommand{\kpr}{k_{\parallel}}
\newcommand{\lpr}{l_{\parallel}}
\newcommand{\ppp}{p_{\perp}}
\newcommand{\lpp}{l_{\perp}}
\newcommand{\kpp}{k_{\perp}}

\date{\today}

\begin{document}

\title{\bf Quantum Kinetic Theory for Quantum Chromodynamics}

\author[]{Shu Lin
\thanks{linshu8@mail.sysu.edu.cn}}
\affil[]{School of Physics and Astronomy, Sun Yat-Sen University, Zhuhai 519082, China}

\maketitle

\begin{abstract}
We develop a quantum kinetic theory for QCD, which incorporates all leading order collision terms. At lowest order in gradient expansion, it reproduces the spin-averaged Boltzmann equation with both elastic and inelastic collisions. At next order in gradient expansion, the solution to the quantum kinetic equations give spin polarization of on-shell quarks and gluons in quark-gluon plasma when the gradients are of hydrodynamic ones. A power counting in the coupling shows the spin polarization behaves differently in vortical and non-vortical gradients: the former is free of collisional contribution to leading order, while the latter contains a collisional contribution at parametrically the same order as the free theory counterpart. We also find the inelastic collision in a spin basis provides a possible mechanism for conversion between spin and orbital angular momentum.
\end{abstract}


\newpage

\section{Introduction}

The past decade has witnessed a growing interest in spin phenomena within the quark-gluon plasma (QGP) created in relativistic heavy-ion collisions. This interest was ignited by the discovery of global polarization of $\Lambda$ hyperons produced in these collisions \cite{STAR:2017ckg,STAR:2018gyt}, confirming early theoretical predictions based on spin-orbit coupling \cite{Liang:2004ph,Liang:2004xn}. Subsequent measurements of local polarization with respect to reaction plane harmonics have unveiled complex vortical structures within the medium \cite{STAR:2019erd,STAR:2023eck}. On the other hand, observations of spin alignment for vector mesons---including $\phi$, $J/\psi$ and $D^{*+}$---have produced puzzling deviations from naive expectations that challenge existing theoretical frameworks~\cite{STAR:2022fan,ALICE:2019aid,ALICE:2022dyy,ALICE:2025cdf}. These experimental discoveries have opened a new a arena in heavy-ion physics: understanding how the spin degrees of freedom (DOF) of quarks and gluons evolve in a rapidly expanding, strongly interacting medium.

One of the widely-used theoretical frameworks for describing such non-equilibrium dynamics is the kinetic theory. At its most basic level, kinetic theory provides a bridge between the microscopic quantum field theory---QCD itself---and the macroscopic observables measured in experiments. The traditional form of the kinetic theory has been the Boltzmann equation, which describes the phase space evolution of quarks and gluons consisting the QGP~\cite{Arnold:2002zm}. This framework has achieved remarkable success in understanding the transport phenomena \cite{Arnold:2000dr,Arnold:2003zc}. However, the Boltzmann equation treats the quarks and gluons as spin averaged quasi-particles, making it inadequate for describing spin DOF.
The necessity of incorporating spin dynamics has driven the development of more sophisticated quantum kinetic theory (QKT). A major progress came with the formulation of QKT for massive fermions, derived from the Wigner-function approach to quantum field theory~\cite{Hattori:2019ahi,Weickgenannt:2019dks,Gao:2019znl,Liu:2020flb,Guo:2020zpa}. Unlike massless Weyl fermions, where spin is enslaved to momentum (helicity), massive fermions possess spin as an independent dynamical degree of freedom. The QKT captures this by introducing coupled kinetic equations for the vector (particle number) and axial (spin) components, see \cite{Hidaka:2022dmn} for a review and references therein. 

While collisionless QKT is well developed and phenomenological interaction based collisional extensions have been written down \cite{Hidaka:2017auj,Zhang:2019xya,Carignano:2019zsh,Yang:2020hri,Wang:2020pej,Shi:2020htn,Weickgenannt:2020aaf,Hou:2020mqp,Yamamoto:2020zrs,Weickgenannt:2021cuo,Sheng:2021kfc,Wang:2021qnt}, a realistic collisional QKT based on QCD is hampered by the complicated QCD interaction. Recently, a QED based QKT has been developed as a prototype of QCD based QKT \cite{Lin:2021mvw}. At lowest order in hydrodynamic gradient, it reduces to the abelian version of the Boltzmann equation, with the collision term shares many similarities with its QCD counterpart, including systematic treatment of screening and medium induced inelastic collision, which is not possible with phenomenological interactions. At next order in gradient, fermion and photon gain spin polarization. In particular, it allows for systematic treatment of collisional contribution to spin polarization \cite{Lin:2022tma,Lin:2024zik,Wang:2024lis,Fang:2023bbw,Fang:2024vds}. Independent of this, QCD based QKT specific for heavy quark and quarkonia have been developed separately for studying the spin evolution of the corresponding DOF in QGP respectively \cite{Li:2019qkf,Hongo:2022izs,Yang:2024ejk,Chen:2025mrf}.

In this paper, we fill the gap by developing the full QCD based QKT. QCD differs from QED in its non-abelian group structure and the gauge particle self-interaction. The former merely introduces trivial modification with appropriate color factors as colored mean field can't exist at long range due to screening effect. The latter introduces more complicated self-energy diagrams accounting for modification of dispersions and collisions. Several new features have been highlighted in the paper: the inelastic collision has interesting interpretation of spectral function of 3-body bound state; while the elastic collision is conveniently discussed in the spin averaged basis, we find it more instructive to discuss the inelastic collision in spin basis, which provides a mechanism of spin and orbital conversion.

The remainder of this paper is organized as follows. In Sec~\ref{DOF}, we briefly discuss the degrees of freedom and hierarchy of scales needed for gradient expansion as preparation for the derivation of QKT. In Sec~\ref{KB} we derive the Kadanoff-Baym equations satisfied by Wigner function of quarks and gluons respectively. in Sec~\ref{se_disp}, we discuss self-energy diagrams at (effective) one loop, which modify of dispersions of the DOF in medium. In Sec~\ref{se_collision}, we discuss two-loop and a specific set of multi-loop diagrams that give rise to elastic and inelastic collisions. Combining the results In Sec~\ref{KB}, \ref{se_disp} and \ref{se_collision}, we reproduce the widely-used Boltzmann equation for QCD. In Sec~\ref{qke_pol}, we derive the QKT at next order in gradient and obtain spin polarizations of quarks and gluons in response to hydrodynamic sources. We distinguish vortical and non-vortical gradients, which contribute to spin polarization in different ways. The QKT is derived within Coulomb gauge. We discuss possible gauge dependencies in Sec~\ref{gauge_dep}. Sec~\ref{sum_out} is devoted to summary and outlook. Calculation details are reserved in one appendix.

\section{Degrees of freedom and gradient expansion}\label{DOF}

The degrees of freedom of the quantum kinetic theory for QCD are on-shell quarks and gluons. The quantum kinetic equations are organized as a gradient expansion. The dimensionless parameter is $\e\sim\hb\pd_X/\L$, with $\pd_X$ and $\L$ being respectively gradient and typical momenta of the DOF. For phenomenological application, one naturally associates the artificial gradient with hydrodynamic ones. For finite temperature QGP, $\L\sim T$. The upper bound of hydrodynamic scale is set by the mean free path $\pd_X\lesssim g^4\L$. At lowest order in gradient (or equivalently $\hb$), we choose the DOF to be unpolarized, i.e. fully described by their spin-averaged distribution functions, and the QKE reduces to the widely-used Boltzmann equation \cite{Arnold:2002zm}. The evolution of spin of the DOF appears at the next order in gradient expansion, for which polarized distribution functions are needed. 

The inverse of the gradient provides a coarse-graining scale for coordinate labeling fluid elements, with dynamics of quarks and gluons localized in each fluid element. While dynamics of on-shell quarks and gluons can induce mean field, they are effectively screened on the coarse-graining scale: quarks and longitudinal gluons are perturbatively screened and transverse gluons are non-perturbatively screened, with screening lengths being $(g\L)^{-1}$ and $(g^2\L)^{-1}$ respectively. Thus we don't need to consider mean field effect. The absence of mean field leads to a nice simplification: the many-body system is color neutral within each fluid element, thus the quarks and gluons have trivial color structure as in vacuum or in thermal equilibrium. It follows that the role of color appears only through Casimir factors, which is the counterpart of charge squared in QED case.

The quarks and gluons gain thermal mass and finite width through interaction with medium, which scale as $g\L$ and $g^2\L$ respectively. The corrections to dispersion relation are subleading for hard on-shell DOF. For elastic collisions, the amplitudes can have infrared (IR) divergences, which occur when soft off-shell particles are exchanged. These IR divergences are cutoff by thermal mass of the soft exchanged particles. For inelastic collisions, these corrections to dispersion relation provide crucial regularization for pinching singularity or collinear divergence.

\section{Kadanoff-Baym Equations for QCD}\label{KB}

The starting point is the Kadanoff-Baym (KB) equation, which can be derived from non-equilibrium field theory on the Schwinger-Keldysh contour \cite{Blaizot:2001nr}. As we discussed in the previous section, the color structure of quarks and gluons is trivial. Thus the formal derivation of the KB equation closely follow the QED case. We shall be brief in the derivation and highlight the key features of QCD case. Readers interested in the details are referred to \cite{Lin:2021mvw}. Below we shall set $\hb=1$. The KB equation for quarks in coordinate space is given by
\begin{align}\label{KB_q_coord}
\(i{\slashed \pd}_x-m\)S^<(x,y)=\int_z\(\S_R(x,z)S^<(z,y)+\S^<(x,z)S_A(z,y)\),
\end{align}
where the lesser/greater propagator defined as
\begin{align}
&S_{\a\b}^<(x,y)=-\<{\bar\ps}_\b(y)\ps_\a(x)\>,\nonumber\\
&S_{\a\b}^>(x,y)=\<\ps_\a(x){\bar\ps}_\b(y)\>.
\end{align}
The lesser self-energy is defined similarly with $\ps\to- g{\slashed A}\ps$ and $\bar{\ps}\to -g\bar{\ps}{\slashed A}$. The expecation value is taken in the medium. We shall also need the retarded/advanced propagators defined as follows
\begin{align}
S_R(x,y)&=i\th(x_0-y_0)\(S^>(x,y)-S^<(x,y)\)\no
S_A(x,y)&=-i\th(y_0-x_0)\(S^>(x,y)-S^<(x,y)\),
\end{align}
and similar for the self-energy. The gradient expansion is achieved with the Wigner transform
\begin{align}\label{Wigner}
\tilde{S}^<(X=\frac{x+y}{2},P)=\int d^4(x-y)e^{iP\cdot (x-y)}\< S^<(x,y)\>,
\end{align}
with $X$ and $P$ being the coarse-grained coordinate and momentum of quark respectively. Performing gradient expansion on \eqref{KB_q_coord}, we obtain
\begin{align}\label{KB_q}
\frac{i}{2}\slashed{\pd}S^<+(\slashed{P}-m)S^<=\(\S_RS^<+\S^<S_A\)+\frac{i}{2}\(\{\S_R,S^<\}+\{\S^<,S_A\}\)_\pb,
\end{align}
where the Poisson bracket is defined as
\begin{align}\label{PB}
\{\tilde{A},\tilde{B}\}_\pb=\pd_k\tilde{A}\cdot\pd_X\tilde{B}-\pd_X\tilde{A}\cdot\pd_k\tilde{B}.
\end{align}
\eqref{KB_q} is to be solved order by order in gradient as
\begin{align}
S^<=S^{<(0)}+S^{<(1)}+\cdots.
\end{align}
We assume $S^{<(0)}(S^{>(0)})$ to have the form of equilibrium propagator in the absence of self-energy
\begin{align}\label{S_less}
&S_{ij}^{<(0)}(X,P)=2\p\e(P\cdot u)\d(P^2-m^2)\({\slashed P}+m\)(-f_q(X,P))\d_{ij},\no
&S_{ij}^{>(0)}(X,P)=2\p\e(P\cdot u)\d(P^2-m^2)\({\slashed P}+m\)(1-f_q(X,P))\d_{ij},
\end{align}
with $f_q(X,P)$ being the unpolarized local quark distribution function. $u(X)$ is the fluid velocity of the local fluid element. The trivial color structure will be suppressed below. Note that $S^{<(0)}$ is nonvanishing only for on shell DOF. As we discussed in the previous section, this is sufficient for describing elastic collisions. For inelastic collisions, we need to keep the medium modification to the dispersion relation. KB equation for retarded propagator is more useful \cite{Lin:2024svh}
\begin{align}\label{KB_q_retarded}
\frac{i}{2}{\slashed \pd}S_R(X,P)+\({\slashed p}-m\)S_R(X,P)-\(\S_R(X,P)S_R(X,P)+\frac{i}{2}\{\S_R(X,P),S_R(X,P)\}_\pb\)=-1.
\end{align}
To lowest order in gradient, we have
\begin{align}\label{S_R0}
S_R^{(0)}(X,P)=\frac{-1}{{\slashed P}-m-\S_R},
\end{align}
with the real and imaginary parts of $\S_R$ providing thermal mass and width respectively. Since \eqref{S_R0} is of local equilibrium form, it is expected to be related to $S^{<(0)}$ as
\begin{align}\label{KMS_q}
S^{<(0)}=-2f_q\text{Im}[S_R^{(0)}].
\end{align}
A similar relation holds for $\S^{<(0)}$ and $\S_R^{(0)}$.
We can easily verify this by rewriting \eqref{KB_q} with following representation \footnote{These relations hold beyond lowest order.}
\begin{align}\label{RA_gl}
&\S_R=\text{Re}\S^R+\frac{i}{2}\(\S^>-\S^<\),\nonumber\\
&S^A=\text{Re}S^R-\frac{i}{2}\(S^>-S^<\).
\end{align}
We obtain the following equation satisfied by $S^{<(0)}$
\begin{align}
\frac{i}{2}\slashed{\pd}S^{<(0)}+(\slashed{P}-m)S^{<(0)}=\text{Re}[\S_R^{(0)}]S^{<(0)}+\S^{<(0)}\text{Re}[S_R^{(0)}]+\frac{i}{2}\(\S^{>(0)}S^{<(0)}-\S^{<(0)}S^{>(0)}\).
\end{align}
Note that by construction $S^{<}$ is hermitian. This allows us to separate the real and imaginary parts of the equation above. The real part gives the dispersion relation. Its consistency with \eqref{KB_q_retarded} can be seen by taking the imaginary part of \eqref{KB_q_retarded} and using \eqref{KMS_q}. The imaginary part gives the Boltzmann equation when proper self-energy is identified. To obtain the Boltzmann equation governing the evolution of $f_q$, we may take the trace of the imaginary part to have
\begin{align}\label{Boltzmann_q}
\tr[{\slashed \pd}S^{<(0)}]=\tr[\S^{>(0)}S^{<(0)}-\S^{<(0)}S^{>(0)}].
\end{align}

The KB equation for gluon is parallel to the counterpart for photon, which is worked out in \cite{Lin:2021mvw,Hattori:2020gqh}. Keeping terms up to $O(\pd)$, we have
\begin{align}\label{KB_g}
&\bigg[-P^2\h^{\m\n}+P^\m P^\n-\frac{1}{\x}P^{\m\a}P^{\n\b}P_\a P_\b+\frac{i}{2}\(-2P\cdot\pd \h^{\m\n}+(\pd^\m P^\n+\pd^\n P^\m)
-\frac{1}{\x}P^{\m\a}P^{\n\b}(\pd_\a P_\b+\pd_\b P_\a)\)\bigg]D^<_{\n\r}\no
&=\text{Re}[\P^{\m\n}_R]D^<_{\n\r}+\P^{\m\n<}\text{Re}[D_{\n\r}^R]+\frac{i}{2}\(\P^{\m\n>}D_{\n\r}^<-\P^{\m\n<}D_{\n\r}^>\)+\frac{1}{4}\(\{\P^{\m\n>},D_{\n\r}^<\}-\{\P^{\m\n<},D_{\n\r}^>\}\).
\end{align}
The lesser/greater propagators are defined as
\begin{align}
&D_{\m\n}^<(X=\frac{x+y}{2},P)=\int d^4(x-y)e^{iP\cdot(x-y)}\<A_\n(y)A_\m(x)\>,\nonumber\\
&D_{\m\n}^>(X=\frac{x+y}{2},P)=\int d^4(x-y)e^{iP\cdot(x-y)}\<A_\m(x)A_\n(y)\>,
\end{align}
and similarly for the self-energies.
$P_{\m\n}=u_\m u_\n-\h_{\m\n}$ is the projector orthogonal to the fluid velocity. There are gauge fixing terms for Coulomb gauge in the KB equation, which follow from the gauge fixing term $-\frac{1}{2\x}\(P^{\a\b}\pd_\a^x A_\b\)^2$. The usual Coulomb gauge is obtained when $\x\to0$. We shall keep $\x$ general in the following
\eqref{KB_g} is also to be solved in gradient expansion as
\begin{align}
D^<=D^{<(0)}+D^{<(1)}+\cdots.
\end{align}
At lowest order, only the transverse gluons are on-shell, thus
\begin{align}\label{D_less}
&D_{\m\n,ab}^{<(0)}(X,P)=2\p\e(p\cdot u)\d(P^2)P_{\m\n}^T f_g(X,P)\d_{ab},\no
&D_{\m\n,ab}^{>(0)}(X,P)=2\p\e(p\cdot u)\d(P^2)P_{\m\n}^T (1+f_g(X,P))\d_{ab},
\end{align}
with $f_g(X,P)$ being the unpolarized local gluon distribution function. The trivial color structure $\d_{ab}$ will be suppressed below. $P_{\m\n}^T=P_{\m\n}-\frac{P_{\m\a}P_{\n\b}P^\a P^\b}{-P^2+(P\cdot u)^2}$ is the transverse projector orthogonal to both the fluid velocity and gluon momentum.
As in the quark case, \eqref{D_less} is sufficient for describing elastic collisions, for which we may drop the self-energy correction to dispersion from $\text{Re}[\P^{\m\n}_R]D^<_{\n\r}+\P^{\m\n<}\text{Re}[D_{\n\r}^R]$ in \eqref{KB_g}. Retaining these terms is necessary for describing inelastic collisions. It is more convenient to use the KB equation for retarded propagator
\begin{align}\label{KB_g_retarded}
\(-P^2g^{\m\n}+P^\m P^\n-\frac{1}{\x}P^{\m\a}P^{\n\b}P_\a P_\b\)D_{\n\r}^{R(0)}(P)-\P^{\m\n{(0)}}_R D_{\n\r}^{R(0)}(P)=-\d^\m_\r.
\end{align}
$\P_{\m\n}^{R(0)}$ can be decomposed as its equilibrium counterpart as
\begin{align}\label{se_Pi}
\P_{\m\n}^R=P_{\m\n}^T\P_T^R-\frac{P^2}{p^2}P_{\m\n}^L\P_L^R,
\end{align}
with $P_{\m\n}^L=-\h_{\m\n}+\frac{P_\m P_\n}{P^2}-P_{\m\n}^T$ being the longitudinal projector. We can then solve $D_{\m\n}^{R(0)}$ as
\begin{align}
D_{\m\n}^{R(0)}(P)=\frac{-1}{P^2-\P_T^{R(0)}}P_{\m\n}^T+\frac{-1}{p^2+\P_L^{R(0)}}u_\m u_\n+\x\frac{P_\m P_\n}{p^4},
\end{align}
with the three terms corresponding to transverse, longitudinal and gauge components. The dispersions of the first two are modified by transverse and longitudinal components of self-energies. The longitudinal component is an emergent DOF due to the medium. It decouples for $P\sim T$ thus is not kept in the kinetic theory. The transverse gluons gain thermal mass and width through real and imaginary parts of $\P_T$ respectively. The transverse components of $D^{<(0)}_{\m\n}$ and $D^{R(0)}_{T,\m\n}=\frac{-1}{P^2-\P_T^{R(0)}}P_{\m\n}^T$ should again be related as
\begin{align}\label{KMS_g}
D^{<(0)}_{\m\n}=2f_g\text{Im}[D^{R(0)}_T]P_{\m\n}^T.
\end{align}
To see this, we plug \eqref{D_less} into \eqref{KB_g} and \eqref{KB_g_retarded} to obtain simplified KB equations for transverse gluons
\begin{align}\label{KB_g_tr}
&\bigg[-P^2\h^{\m\n}+\frac{i}{2}\(-2P\cdot\pd \h^{\m\n}+\pd^\n P^\m
-\frac{1}{\x}P^{\m\a}P^{\n\b}\pd_\b P_\a\)\bigg]D^<_{\n\r(0)}
=\text{Re}[\P^{T(0)}_R]P^{\m\n}_TD^{<(0)}_{\n\r}+\P^{<(0)}_TP^{\m\n}_T\text{Re}[D_{\n\r}^{R(0)}] \no
&+\frac{i}{2}\(\P^{\m\n>(0)}D_{\n\r}^{<(0)}-\P^{\m\n<(0)}D_{\n\r}^{>(0)}\)+\frac{1}{4}\{\P^{\m\n>(0)},D_{\n\r}^{<(0)}\}
+\text{P.B.}.
\end{align}
The counterpart for $D_{T,\n\r}^{R(0)}$ can be deduced from \eqref{KB_g_retarded} as
\begin{align}\label{KB_g_retarded_tr}
&\(-P^2\h^{\m\n}\)D_{T,\n\r}^{R(0)}(P)-\P^{R{(0)}}_TP^{\m\n}_TD^{R(0)}_{T,\n\r}=-\d^\m_\r.
\end{align}
As in the quark case, we can immediately verify the consistency between the real part of \eqref{KB_g_tr} and the imaginary part of \eqref{KB_g_retarded_tr} using representation similar to \eqref{RA_gl} and \eqref{KMS_g}.
Similar to the quark case, the imaginary part of \eqref{KB_g_tr} gives the Boltzmann equation. Note that \eqref{KB_g_tr} is not completely transverse as the index $\m$ can be longitudinal. We may further project its imaginary part by contracting it with $P_\m^{\r,T}$ to arrive at the following simple form
\begin{align}\label{Boltzmann_g}
-2P\cdot\pd P^{\n\r}_T D_{\n\r}^{<(0)}=\P^{\m\n>(0)}D_{\n\m}^{<(0)}-\P^{\m\n<(0)}D_{\n\m}^{>(0)}.
\end{align}
The color structure of the kinetic equations can be restored by noting that both propagators and self-energies are diagonal in color, thus explicit color contractions are given by
\begin{align}
&\S^>_{ik}S^<_{kj}\sim \d_{ik}\d_{kj}=\d_{ij},\no
&\P^{\m\n>}D_{\n\m}^<\sim \d_{ac}\d_{cb}=\d_{ab},
\end{align}
for quarks and gluons respectively. It is convenient to perform color sums on both sides of kinetic equations \eqref{Boltzmann_q} and \eqref{Boltzmann_g} to have
\begin{align}
&d_F{\slashed \pd}S^{<(0)}=\Tr[\S^{>(0)}S^{<(0)}-\S^{<(0)}S^{>(0)}],\no
&-2d_A P\cdot\pd P^{\n\r}_T D_{\n\r}^{<(0)}=\Tr[\P^{\m\n>(0)}D_{\n\m}^{<(0)}-\P^{\m\n<(0)}D_{\n\m}^{>(0)}],
\end{align}
where $d_F=N_c$ and $d_A=N_c^2-1$. $\Tr$ is defined as trace over color (and spin when needed). For gluons, the sum over spin is explicit in the contraction of Lorentz indices.

\section{Self-energies and modified dispersions}\label{se_disp}

\subsection{Thermal mass}
Now we work out medium modification to the dispersions from $\S_R^{(0)}$ and $\P_R^{T,(0)}$ for quarks and gluons respectively, which can be calculated with the propagators $S_R^{(0)}$ and $D^{R(0)}_{\m\n}$. We shall suppress the superscript $(0)$ for simplicity. We proceed in the $ra$-basis with $\S_R=-i\S_{ar}$ and $\P_R^{\m\n}=-i\P_{ar}^{\m\n}$. The quark propagators in $ra$-basis are given by
\begin{align}\label{ra_prop_q}
&S_{ra}(Q)=\frac{i({\slashed Q}+m)}{(q_0+i\h)^2-q^2-m^2},\nonumber\\
&S_{ar}(Q)=S_{ra}(\h\to-\h),\nonumber\\
&S_{rr}(Q)=-2\p\e(q_0)\d(Q^2-m^2)({\slashed Q}+m)f_q(X,Q).
\end{align}
The gluon propagators in Coulomb gauge are given by
\begin{align}\label{ra_prop_g}
&D_{\m\n}^{ra}(Q)=\frac{i}{(q_0+i\h)^2-\vec{q}^2}\(P_{\m\n}^T+\frac{Q^2u_\m u_\n}{q^2}\),\nonumber\\
&D_{\m\n}^{ar}(Q)=\frac{i}{(q_0-i\h)^2-\vec{q}^2}\(P_{\m\n}^T+\frac{Q^2u_\m u_\n}{q^2}\),\nonumber\\
&D_{\n\m}^{rr}(Q)=2\p\e(q_0)\d(Q^2)P_{\m\n}^T f_g(X,Q).
\end{align}
$f_{q/g}(X,Q)$ are the local equilibrium distributions. As we shall see, the one-loop self-energies are local in the sense that they depend on integration of $f_{q/g}(X,Q)$ only. The vertices are the same as bare ones. We will need self-energy for hard on-shell quarks/gluons and soft off-shell quarks/gluons. For the latter case, thermal mass and width are captured by the self-energies in the HTL approximation. The former case necessarily requires going beyond the HTL limit. The former case will be attained by power corrections, which are in general gauge dependent. We will comment on the effect of power corrections based on existing results in literature.

\begin{figure}
	\includegraphics[width=0.45\textwidth]{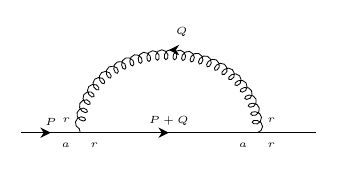}
	\includegraphics[width=0.45\textwidth]{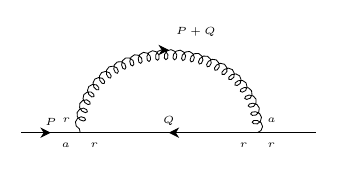}
	\caption{\label{quark_se}Retarded quark self-energy $\S_{ar}$ diagrams in $ra$-basis.}
\end{figure}
\begin{figure}
	\includegraphics[width=0.45\textwidth]{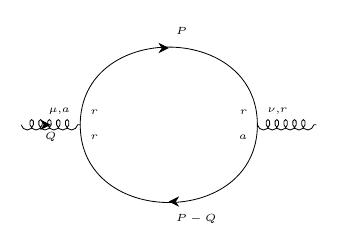}
	\includegraphics[width=0.45\textwidth]{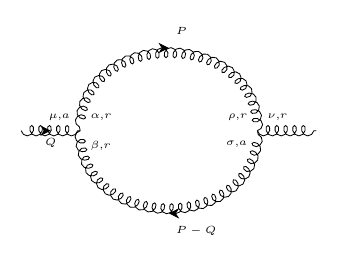}
	\includegraphics[width=0.45\textwidth]{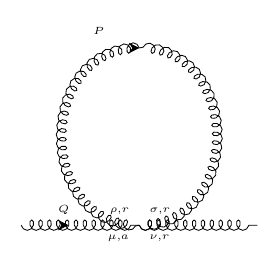}
	\caption{\label{gluon_se}Retarded gluon self-energy $\P_{ar}^{\m\n}$ diagrams in $ra$-basis. For the first two diagrams, an inequivalent diagram can be obtained by swaping the $ra$ labelings of the upper and lower propagators in the loop. Color indices have been suppressed for clarity.}
\end{figure}

The leading one-loop diagrams are shown in and Fig.~\ref{quark_se} and Fig.~\ref{gluon_se} for quark and gluon respectively. We first calculate the quark self-energy. The left diagram is given by
\begin{align}\label{Sigma_left}
\S^{ar}_1(P)=-g^2C_F\int_Q\frac{i\g^\n({\slashed P}+{\slashed Q}+m)\g^\m}{(p_0+q_0+i\h)^2-(\vec p+\vec q)^2-m^2}P_{\m\n}^T(Q)\(\frac{1}{2}+f_g(q_0)\)2\p\e(q_0)\d(Q^2),
\end{align}
where the color factor comes from $t^a_{ik}t^a_{kj}=C_F\d_{ij}$ and $\int_Q\equiv\int\frac{d^4Q}{(2\p)^4}$.
We will need the self-energy for hard on-shell quark or soft off-shell one, for which we may require $m\lesssim g\L$ such that the inelastic collision is relevant and the screening of exchanged quark is necessary. This allows us to drop $m$ in the numerator. Since $P$ can be either hard on-shell or soft off-shell, we may also drop the $i\h$ and $m^2$ in the denominator. Moreover, the term proportional to ${\slashed P}$ is odd in $Q$ thus integrated to zero. The remaining product of gamma matrices are evaluated as
\begin{align}\label{product_gamma}
\g^\n{\slashed Q}\g^\m P_{\m\n}^T(Q)=\(Q^\n\g^\m-{\slashed Q}\h^{\m\n}+Q^\m\g^\n\)P_{\m\n}^T(Q)=2{\slashed Q}.
\end{align}
Note that we would obtain the same if we use  $P_{\m\n}^T\to-\h_{\m\n}$, corresponding to the Feynman gauge. This shows that the unphysical polarizations don't contribute. We then have
\begin{align}
\S^{ar}_1(P)=\int_Q\frac{i{\slashed Q}}{P\cdot Q}f_g(q_0)2\p\e(q_0)\d(Q^2),
\end{align}
where the factor $1/2$ corresponding to the vacuum contribution has been dropped.
The right diagram is evaluated similarly as
\begin{align}\label{Sigma_right}
&\S^{ar}_2(P)=-ig^2C_F\int_Q\frac{\g^\n(-{\slashed Q}+m)\g^\m}{(P+Q)^2}\(P_{\m\n}^T(P+Q)+u_\m u_\n\frac{(P+Q)^2}{({\vec p}+\vec{q})^2}\)\(\frac{1}{2}-f_q(q_0)\)2\p\e(q_0)\d(Q^2-m^2).
\end{align}
The product of gamma matrices is evaluated using \eqref{product_gamma}. $(P+Q)^2=P^2+2P\cdot Q+m^2\simeq 2P\cdot Q$, with the equality enforced by $\d(Q^2-m^2)$ and the approximation holds for our kinematic restrictions on $P$ and $m$. The term proportional to $u_\m u_\n$ from longitudinal gluon is suppressed as $\frac{(P+Q)^2}{({\vec p}+\vec{q})^2}\simeq\frac{2P\cdot Q}{q^2} \ll1$. In principle there could be power correction for hard $P$. This has been considered in \cite{Carignano:2017ovz} for QED. The results are gauge dependent and generalize straightforwardly to QCD. The power correction doesn't give rise to thermal mass. Thus we find
\begin{align}
\S^{ar}_2(P)\simeq\int_Q\frac{-{\slashed Q}}{P\cdot Q}\(-f_q(q_0)\)2\p\e(q_0)\d(Q^2).
\end{align}

Next we turn to gluon self-energy. The quark loop diagram gives the following contribution to self-energy as
\begin{align}\label{Pi_1}
&\P_1^{ar}(Q)=2g^2C_F N_f\int_Q\tr[\g^\n({\slashed P}+m)\g^\m({\slashed P}+{\slashed Q}+m)]\frac{i}{(P-Q)^2-m^2}\d(P^2-m^2)2\p\e(p_0)(\frac{1}{2}-f_q(p_0)).
\end{align}
The color factor comes from $tr[t^at^b]=\d_{ab}C_F$. A flavor factor $N_f$ has been inserted. The factor of $2$ includes an identical contribution from the other quark loop diagram from a different $ra$ labelings described in the Fig.~\ref{gluon_se}.
The trace is evaluated as
\begin{align}\label{trace}
\tr[\g^\n({\slashed P}+m)\g^\m({\slashed P}+{\slashed Q}+m)]\simeq 8P^\m P^\n-4(P^\m Q^\n+P^\n Q^\m)+4\h^{\m\n}P\cdot Q,
\end{align}
with $m$ dependence ignored as in the quark case. Using the on-shell condition $\d(P^2-m^2)$, we can simplify the denominator as
\begin{align}
\frac{1}{(P-Q)^2-m^2}=\frac{1}{-2P\cdot Q+Q^2}\simeq\frac{1}{-2P\cdot Q}\(1+\frac{Q^2}{2P\cdot Q}\),
\end{align}
which can be combined with the trace as
\begin{align}
&\tr[\g^\n({\slashed P}+m)\g^\m({\slashed P}+{\slashed Q}+m)]\frac{1}{(P-Q)^2-m^2}\no
&\simeq\[\frac{8P^\m P^\n}{-2P\cdot Q}+\frac{-4(P^\m Q^\n+P^\n Q^\m)+4\h^{\m\n}P\cdot Q}{-2P\cdot Q}-\frac{8P^\m P^\n Q^2}{(2P\cdot Q)^2}+\cdots\].
\end{align}
The first term integrates to zero as the remainder of the integrand is odd in $Q$. The other two terms give
\begin{align}\label{ql_se}
\P_1^{ar}(Q)=2ig^2C_FN_f\int_Q\[\frac{-4(P^\m Q^\n+P^\n Q^\m)+4\h^{\m\n}P\cdot Q}{-2P\cdot Q}-\frac{8P^\m P^\n Q^2}{(2P\cdot Q)^2}\]2\p\e(p_0)(-f_q(p_0)).
\end{align}
The $\cdots$ term gives power correction to gluon self-energy. QED analog of this contribution has been found in \cite{Gorda:2022fci}, with the power correction not contributing to medium induced mass for hard on-shell momentum.

The gluon loop diagram involving 3-gluon vertices is given by
\begin{align}\label{Pi_3g}
&\P_2^{ar}(Q)=g^2C_A\int_P P_{\a\r}^T(P)\(P_{\b\s}^T(P-Q)+u_\b u_\s\frac{(P-Q)^2}{({\vec p}-{\vec q})^2}\)\frac{i}{(P-Q)^2}2\p\e(p_0)\(\frac{1}{2}+f_g(p_0)\)\times \no
& \[\h^{\m\a}(Q+P)^\b+\h^{\a\b}(-2P+Q)^\m+\h^{\b\m}(P-2Q)^\a\]\[\h^{\n\r}(-Q-P)^\s+\h^{\r\s}(2P-Q)^\n+\h^{\s\n}(-P+2Q)^\r\].
\end{align}
The color factor comes from $f_{acd}f_{bcd}=C_A\d_{ab}$. The symmetry factor $1/2$ is compensated by identical contributions from two diagrams with different $ra$ labelings.
Note that there are both transverse and longitudinal contributions for the off-shell gluon in the loop. We calculate them separately below. For the transverse contribution, we use the property of transverse projector and $\d(P^2)$ to simplify the product of square brackets as
\begin{align}\label{square_tt}
&\[\cdots\]\[\cdots\]P_{\a\r}^T(P)P_{\b\s}^T(P-Q)\no
&=
\[\h^{\m\a}2Q^\b+\h^{\a\b}(-2P+Q)^\m-\h^{\b\m}(-2Q^\a)\]\[-\h^{\n\r}2Q^\s+\h^{\r\s}(2P-Q)^\n+\h^{\s\n}2Q^\r\]P_{\a\r}^T(P)P_{\b\s}^T(P-Q)\no
&\simeq\[-8P^\m P^\n+4P^\m Q^\n+4P^\n Q^\m\],
\end{align}
where we have approximated $P_{\b\s}^T(P-Q)\simeq P_{\b\s}^T(P)$ and keep up to next to leading order in $P$. For the longitudinal component, we first approximate the projector as $u_\b u_\s\frac{(P-Q)^2}{({\vec p}-{\vec q})^2}\simeq u_\b u_\s\frac{-2P\cdot Q}{p^2}$ using $\d(P^2)$. Note that this is subleading compared to the transverse counterpart, thus we only need to keep the leading contribution in the product of square brackets to obtain
\begin{align}\label{square_tl}
&\[\cdots\]\[\cdots\]P_{\a\r}^T(P) u_\b u_\s\frac{-2P\cdot Q}{p^2}\no
&\simeq-P^{\m\n}_T(P)\frac{-2P\cdot Q}{p^2}p_0^2.
\end{align}
Similar to the quark loop case, we need to expand the denominator as
\begin{align}
\frac{i}{(P-Q)^2}\simeq\frac{i}{-2P\cdot Q}\(1+\frac{Q^2}{2P\cdot Q}\)
\end{align}
Mutipling to \eqref{square_tt} and \eqref{square_tl}, we obtain
\begin{align}\label{Pi_2}
\P_2^{ar}(Q)=ig^2C_A\int_P\(\frac{4(P^\m Q^\n+P^\n Q^\m)}{-2P\cdot Q}+\frac{-8P^\m P^\n}{-(2P\cdot Q)^2}-P_T^{\m\n}(P)\)2\p\e(p_0)f_g(p_0).
\end{align}
The last diagram is calculated as
\begin{align}\label{Pi_3}
&\P_3^{ar}(Q)=-\frac{1}{2}ig^2C_A\int_P\[\(\h^{\m\n}\h^{\r\s}-\h^{\m\s}\h^{\n\r}\)+\(\h^{\m\n}\h^{\r\s}-\h^{\m\r}\h^{\n\s}\)\]P_{\r\s}^T(P)2\p\e(p_0)f_g(p_0)\no
&\simeq-ig^2C_A\int_P\[-2\h^{\m\n}-P_T^{\m\n}(P)\]2\p\e(p_0)f_g(p_0).
\end{align}
Adding \eqref{ql_se}, \eqref{Pi_2} and \eqref{Pi_3}, we obtain
\begin{align}
&\P_1^{ar}(Q)+\P_2^{ar}(Q)+\P_3^{ar}(Q)\simeq 2ig^2C_FN_f\int_P\(\frac{2(P^\m Q^\n+P^\n Q^\m)}{-P\cdot Q}+\frac{2P^\m P^\n Q^2}{(P\cdot Q)^2}+2\h^{\m\n}\)2\p\e(p_0)f_q(p_0)\no
&+ig^2C_A\int_P\(\frac{2(P^\m Q^\n+P^\n Q^\m)}{-P\cdot Q}+\frac{2P^\m P^\n Q^2}{(P\cdot Q)^2}+2\h^{\m\n}\)2\p\e(p_0)f_g(p_0).
\end{align}
Power corrections to the self-energy have been considered in \cite{Gorda:2023zwy}. It is found that for the on-shell transverse gluons, the power corrections vanish identically in Feynman gauge.

\subsection{Thermal width}
\begin{figure}
	\includegraphics[width=0.45\textwidth]{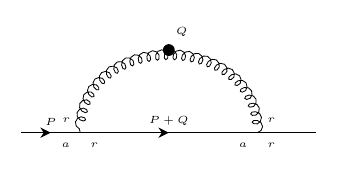}
	\includegraphics[width=0.45\textwidth]{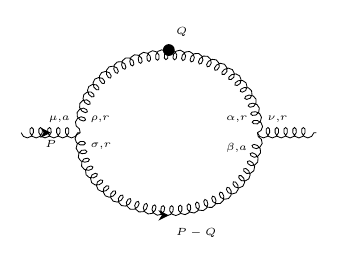}
	\caption{\label{widths}Effective one-loop diagrams contributing to thermal width of quark and gluon. The heavy dotted gluon propagator corresponds to HTL resummed one. For the second diagram, an inequivalent diagram can be obtained by swaping the $ra$ labelings of the upper and lower propagators in the loop.}
\end{figure}

For our purpose, we only need the thermal width for hard on-shell quarks and gluons. These quantities are known to be IR divergent. We aim at an integral representation of the width. The divergence will be cutoff properly in inelastic collision terms. The thermal width can be identified in the retarded propagators of quarks and transverse gluons as
\begin{align}
&\G^q(P)=-\frac{1}{2p_0}\text{Re}\(\tr[({\slashed P}+m)]\S_{ar}(P)\),\no
&\G^g(P)=-\frac{1}{p_0}\text{Re}[\P^{ar}_T],
\end{align}
with $\G_p^q$ and $\G_p^g$ being thermal width for quarks and gluons respectively. Note that the one-loop contribution to self-energy in the previous subsection is purely imaginary, thus can't contribute to thermal width. The leading contribution to the width comes from the diagram in Fig.~\ref{widths}. It involves a soft gluon in the loop \cite{Pisarski:1993rf}. The quark self-energy diagram is evaluated
\begin{align}
&\S_{ar}(P)=g^2C_F\int_Q\frac{i\g^\n({\slashed P}+{\slashed Q}+m)\g^\m}{(P+Q)^2-m^2+i\e(p_0+q_0)\h}D_{rr,\m\n}^*(Q).
\end{align}
with $D_{rr,\m\n}^*$ being the resummed soft gluon propagator with $Q\ll P$. With the kinematic restriction, we can easily take the real part as
\begin{align}
&\text{Re}\[\frac{i}{(P+Q)^2+i\e(p_0-q_0)\h-m^2}\]\simeq\text{Re}\[\frac{i}{-2P\cdot Q+i\e(p_0)\h}\]\no
&=\p\e(p_0)\d(2P\cdot Q)=\frac{2\p}{4p_0}\d(q_0-q_\pr),
\end{align}
where on-shell condition for $P$ has been used. $q_\pr={\vec q}\cdot\he$ and $\he$ is a unit vector almost parallel to ${\vec p}$, with the relative angle between $\he$ and $\vec{p}$ being $O(g)$. The purpose of this choice will be clear soon.
We can also drop the ${\slashed Q}$ in the trace to have
\begin{align}
\tr[({\slashed P}+m)\g^\n({\slashed P}+{\slashed Q}+m)\g^\m]\simeq 4(2P^\m P^\n-(P^2-m^2)\h^{\m\n})=8P^\m P^\n,
\end{align}
where we have used the on-shell condition for $P$. We have then
\begin{align}\label{Gamma_q}
&\G^q(P)=-g^2C_F\frac{1}{2p_0}\int_Q\frac{2\p}{4p_0}\d(q_0-q_\pr)8P^\m P^\n D_{rr,\m\n}^*(Q)\no
&\simeq-g^2C_F\int_Q 2\p\d(q_0-q_\pr)\he^\m\he^\n D_{rr,\m\n}^*(Q).
\end{align}

The gluon self-energy diagram in Fig.~\ref{widths} is given by
\begin{align}
&\P_{ar}^{\m\n}(P)=-g^2C_A\int_Q D^*_{rr,\r\a}(Q)D_{ra,\s\b}(P-Q)\[\h^{\m\r}(P+Q)^\s+\h^{\r\s}(-2Q+P)^\m+\h^{\s\m}(Q-2P)^\r\]\no
&\times
\[\h^{\n\a}(-P-Q)^\b+\h^{\a\b}(2Q-P)^\n+\h^{\b\n}(-Q+2P)^\a\].
\end{align}
The symmetry factor $1/2$ is again compensated by identical contributions from two diagrams with different $ra$ labelings. $D_{rr,\r\a}^*$ denotes the resummed propagator with $Q\ll P$. With the kinematic restrictions, we have
\begin{align}
&D_{ra,\s\b}(P-Q)=\frac{i}{(P-Q)^2+i\e(p_0-q_0)\h}\(P_{\s\b}^T(P-Q)+\frac{(P-Q)^2}{({\vec p}-{\vec q})^2}u_\s u_\b\)\no
&\simeq\frac{i}{(P-Q)^2+i\e(p_0-q_0)\h}\(P_{\s\b}^T(P)+\frac{-2P\cdot Q}{p^2}u_\s u_\b\),
\end{align}
where we have used on-shell conditions for $P$ and $Q$. Clearly the longitudinal component is subleading compared to the transverse component, thus will be dropped. We can then simplify the product of square brackets as
\begin{align}\label{product_width}
&\[\h^{\m\r}(P+Q)^\s+\h^{\r\s}(-2Q+P)^\m+\h^{\s\m}(Q-2P)^\r\]\[\h^{\n\a}(-P-Q)^\b+\h^{\a\b}(2Q-P)^\n+\h^{\b\n}(-Q+2P)^\a\]P_{\s\b}^T(P)\no
&\simeq-4P_T^{\m\n}P^\m P^\a\simeq-4P_T^{\m\n}p_\pr^2\he^\r\he^\a,
\end{align}
with $p_\pr={\vec p}\cdot\he$. We have also defined $e^\m=(1,\he)$ and used $P^\m\pr e^\m$ in the last line. \eqref{product_width} also shows that the resulting self-energy is automatically transverse.
The real part is then taken similarly to give
\begin{align}
&\text{Re}\[\frac{i}{(P-Q)^2+i\e(p_0-q_0)\h}\]\simeq\frac{2\p}{4p_0}\d(q_0-q_\pr),
\end{align}
We then have
\begin{align}\label{Gamma_g}
&\G^g(P)=-\frac{1}{p_0}\text{Re}[\P_{ar}^T]=g^2C_A\frac{4p_\pr^2}{p_0}\int_Q\he^{\r}\he^\a D_{rr,\r\a}^*(Q)\frac{2\p}{4p_0}\d(q_0-q_\pr)\no
&\simeq g^2C_A\int_Q\he^{\r}\he^\a D_{rr,\r\a}^*(Q)2\p\d(q_0-q_\pr),
\end{align}
where on-shell condition $\d(P^2)$ has been used. \eqref{Gamma_q} and \eqref{Gamma_g} show that the widths are independent of particle momentum.

\section{Self-energies and collision terms}\label{se_collision}

We have already encountered two types of self-energy diagrams. One-loop diagrams with only hard particles in the loop and one-loop diagrams with one soft particle in the loop. The particles can't be both on-shell for simple kinematic reasons, thus these self-energy diagrams can't be interpreted as collision term in kinetic theory. The collision terms necessarily arise beyond one-loop diagrams. The leading contributions come from two-loop diagrams and a set of enhanced multi-loop diagrams corresponding to elastic and inelastic collisions respectively. Despite of formal suppression of the collision terms by $O(g^4)$, they can lead to significant deviation of the local equilibrium state. In the presence of hydrodynamic source, the many-body system will be driven to a steady state, with the local distribution deviating from the equilibrium one by $O(\pd/g^4)$. Such deviation is difficult to attain in diagrammatic approach \cite{Gagnon:2006hi,Gagnon:2007qt}, but is efficiently described by kinetic approach \cite{Arnold:2000dr,Arnold:2003zc}.

\subsection{elastic collision}

We first analyze two-loop self-energy diagrams in QCD. Many of them have counterparts in QED, which is analyzed in detail in \cite{Lin:2021mvw}. We shall not elaborate on those diagrams but only focus on diagrams specific to QCD. For the purpose of connecting to kinetic theory, we adopt a slightly different convention from usual Feynman diagrams: we only keep momentum flow on propagators, and assign opposite momentum flow direction according to the sign of energy, i.e. a single momentum flow in Feynman diagram will be split into two in our case, which corresponds to positive and negative energies.

\begin{figure}
	\includegraphics[width=0.45\textwidth]{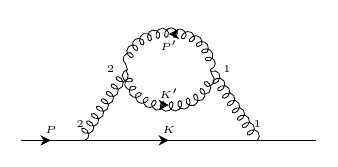}
	\caption{\label{q_Comp_prop}QCD specific quark self-energy diagram of propagator correction type. $1$ and $2$ label position of vertices on the Schwinger-Keldysh contour. Arrows indicate momentum flow, see text for more details (same below).}
\end{figure}
\begin{figure}
	\includegraphics[width=0.45\textwidth]{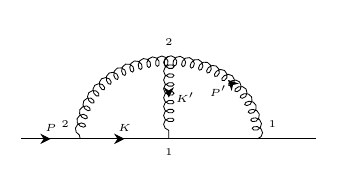}
	\includegraphics[width=0.45\textwidth]{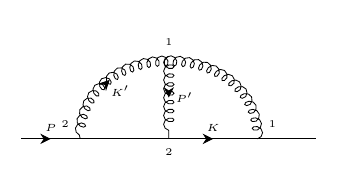}
	\caption{\label{q_Comp_vert}QCD specific quark self-energy diagram of vertex correction type. There are two possible contour labelings. The dummy momentum variables have been relabeled for convenience.}
\end{figure}
We start with quark self-energy diagrams. QCD specific diagram include Fig.~\ref{q_Comp_prop} for propagator corrections and Fig.~\ref{q_Comp_vert} for vertex corrections. Now we show Fig.~\ref{q_Comp_prop} corresponds to t-channel amplitude square of Compton scattering, to be abbreviated as $|t|^2$, and Fig.~\ref{q_Comp_vert} give rise to the interference contributions $ut^*$ and $t s^*$. We start with Fig.~\ref{q_Comp_prop}. To save notations, we will suppress the superscript $(0)$ in this section. Using \eqref{S_less} and \eqref{D_less}, $\tr[\S^>(P)S^<(P)]$ is given by
\begin{align}\label{3g_Sigma}
&\Tr[\S^>(P)S^<(P)]=\frac{1}{2}\int_{P',K'}
\tr[\g^\n({\slashed K}+m)\g^\m({\slashed P}+m)]D_{22,\m\r}(Q)D_{11,\n\s}(Q)P^T_{\a\g}(P')P^T_{\b\d}(K')\times\no
&V^{\r\a\b}(Q,P',-K')V^{\s\g\d}(-Q,-P',K')(2\p)^4\th(p_0)\th(p_0')\th(k_0)\th(k_0')(-f_q(p_0))(1-f_q(k_0))f_g(p_0')(1+f_g(k_0')),
\end{align}
with $V^{\r\a\g}(Q,P',-K')$ and $V^{\s\b\d}(-Q,-P',K')$ being two 3-gluon vertices. All color indices are to be summed over and not shown explicitly.
We have replaced the sign function by heaviside theta function based on the momentum direction labels in the diagram. We will use the abbreviation $\int_{P.S.}=\int_{P',K'}(2\p)^4\th(p_0)\th(p_0')\th(k_0)\th(k_0')(-f_q(p_0))(1-f_q(k_0))f_g(p_0')(1+f_g(k_0'))$ for phase space integration.
Using the following representation
\begin{align}\label{spin_sum}
&\slashed{P}+m=\sum_s u_s(P)\bar{u}_s(P),\no
&P_{\n\m}^T(P')=\sum_i\e_\n^i(P')\e_\m^{i*}(P'),
\end{align}
we can rewrite \eqref{3g_Sigma} as
\begin{align}
&\Tr[\S^>(P)S^<(P)]=\frac{1}{2}\int_{P.S.}\sum_{s,t,i,j}\bar{u}_s(K)\g^\m u_t(P)\bar{u}_t(P)\g^\n u_s(K)\e_\a(P')^i\e_\g^{i*}(P')\e_\b^{j*}(K')\e_\d^{j}(K')\times\no
&V^{\r\a\b}(Q,P',-K')V^{\s\g\d}(-Q,-P',K')\no
&=\frac{1}{2}\int_{P.S.}\sum_{\text{spin,color}}|{\cal T}(P,K,P',K')|^2,
\end{align}
with ${\cal T}(P,K,P',K')$ being t-channel amplitude of Compton scattering. Thus the diagram gives half of $|t|^2$ summed over initial and final spin states integrated over phase space. The other half is given by an inequivalent diagram with momentum direction of $P'$ and $K'$ flipped. Flipping the momentum direction of $P$ and $K$ would lead to counterpart for Compton scattering between anti-quarks and gluons.

Next we turn to Fig.~\ref{q_Comp_vert}. We obtain the following for the left diagram
\begin{align}
&\Tr[\S^>(P)S^<(P)]=\int_{P',K'}
\tr[\g^\n S_{11,\n\s}(P+P')\g^\s({\slashed K}+m)\g^\m({\slashed P}+m)]D_{22,\m\r}(P-K)P^T_{\a\s}(K)P^T_{\b\n}(P')\times\no
&V^{\r\a\b}(P-K,-K',P')(2\p)^4\th(p_0)\th(p_0')\th(k_0)\th(k_0')(-f_q(p_0))(1-f_q(k_0))f_g(p_0')(1+f_g(k_0')),
\end{align}
Using \eqref{spin_sum}, we can rewrite it as
\begin{align}
&\Tr[\S^>(P)S^<(P)]=\int_{P.S.}
\sum_{s,t,i,j}\bar{u}_t(P)\g^\n S_{11}(P+P')\g^\s u_s(K) \bar{u}_s(K)\g^\m u_t(P) D_{22,\m\r}(P-K)\times\no
&\e_\a^{i*}(K')\e_\s^{i}(K)\e_\b^j(P')\e_\n^{j*}(P')
V^{\r\a\b}(P-K,-K',P')\no
&=\int_{P.S.}
\sum_{\text{spin,color}}{\cal S}^*(P,K,P',K'){\cal T}(P,K,P',K'),
\end{align}
with ${\cal S}$ being the s-channel amplitude for Compton scattering. This corresponds the interference term $s^*t$. The right diagram is analyzed similarly to give $u t^*$. There are two more inequivalent diagrams, which can be obtained by $P'\leftrightarrow -K'$ and simultaneous reversing the momentum direction. The corresponding diagrams give the complex conjugates $st^*$ and $u^*t$. There also diagrams with QED analogs, which have been shown to give $|s+u|^2$ \cite{Lin:2021mvw}. Collecting everything, we arrive at the full Compton scattering amplitude square
\begin{align}
\Tr[\S^>(P)S^<(P)]=\int_{P.S.}
\sum_{\text{spin,color}}|{\cal S}(P,K,P',K')+{\cal T}(P,K,P',K')+{\cal U}(P,K,P',K')|^2.
\end{align}
From the phase space measure, we clearly identify it as the loss term for quark. The other term $\tr[\S^<(P)S^>(P)]$ gives the gain term. The processes of Coulomb scattering and pair annihilation don't contain diagram unique to QCD, thus can be easily deduced from QED counterparts.
The diagrams corresponding to Coulomb scattering between quarks are completely analogous to QED. We shall not elaborate on the calculations here.
\begin{figure}
	\includegraphics[width=0.45\textwidth]{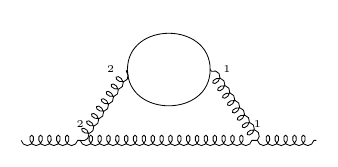}
	\caption{\label{g_Comp_prop}QCD specific gluon self-energy diagram of propagator correction type.}
\end{figure}
\begin{figure}
	\includegraphics[width=0.45\textwidth]{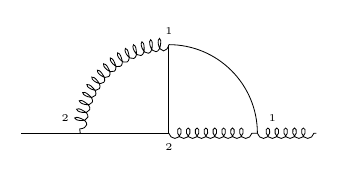}
	\includegraphics[width=0.45\textwidth]{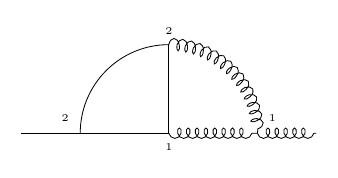}
	\caption{\label{g_Comp_vert}QCD specific gluon self-energy diagram of vertex correction type. There are two possible contour labelings as in the quark case.}
\end{figure}

The gluon self-energy is more complicated. We first consider diagrams corresponding to Compton scattering. The QCD specific diagrams are Fig.~\ref{g_Comp_prop} for propagator correction and Fig.~\ref{g_Comp_vert} for vertex correction. In fact, these diagrams are closely related to their counterparts in quark self-energy. To see that, we note $\Tr[\S^>(P)S^<(P)]$ and $\Tr[\P^{>\m\n}D_{\n\m}^<]$ amounts to joining the external lines of the self-energy diagrams, which give identical diagrams up to relabeling of momentum. The only difference is that the momentum of the joined propagator is not to be integrated over. This simply gives different phase spaces. Since the product of amplitudes are given by unintegrated diagrams, we find the diagrams in Fig.~\ref{g_Comp_prop} and Fig.~\ref{g_Comp_vert} give the same product of amplitudes as the quark case: $|{\cal T}|^2$, ${\cal T}^*({\cal S}+{\cal U})+c.c.$. Combining these with contribution from diagrams having QED analogs, we obtain 
\begin{align}
\Tr[\P^{\m\n>}(P')D_{\n\m}^<(P')]=\int_{P.S.}
\sum_{\text{spin,color}}|{\cal S}(P,K,P',K')+{\cal T}(P,K,P',K')+{\cal U}(P,K,P',K')|^2.
\end{align}
\begin{figure}
	\includegraphics[width=0.45\textwidth]{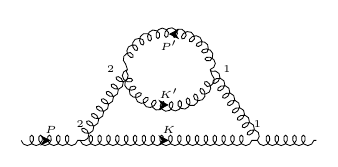}
	\includegraphics[width=0.45\textwidth]{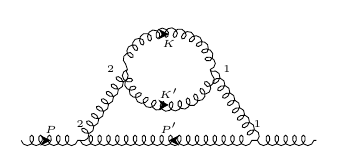}
	\caption{\label{g_Coul_prop}Gluon self-energy diagrams of the propagator correction type with 3-gluon vertices only.}
\end{figure}
\begin{figure}
	\includegraphics[width=0.45\textwidth]{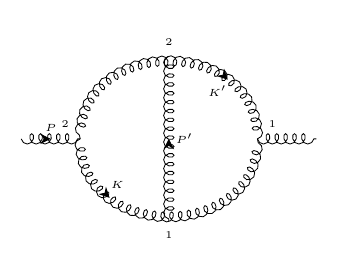}
	\includegraphics[width=0.45\textwidth]{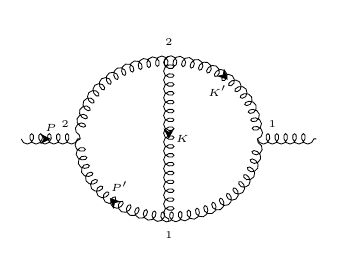}
	\includegraphics[width=0.45\textwidth]{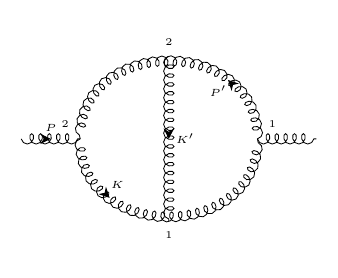}
	\caption{\label{g_Coul_vert}Gluon self-energy diagrams of the vertex correction type with 3-gluon vertices only.}
\end{figure}
Now we turn to Coulomb scatterings, which involves both 3-gluon and 4-gluon vertices. There is no counterpart in QED, with the self-energy diagrams in Fig.~\ref{g_Coul_prop} for propagator correction and Fig.~\ref{g_Coul_vert} for vertex correction. The first type has three possible momentum direction labelings. We give explicit expression for the left diagram of Fig.~\ref{g_Coul_prop}
\begin{align}
&\Tr[\P^{\m\n>}(P)D_{\n\m}^<(P)]=\frac{1}{2}2\int_{P.S.}P^{\a\g}_T(K)P^{\h\l}_T(K')P^{\t\k}_T(P')P^{\n\m}_T(P)D_{22}^{\b\r}(Q)D_{11}^{\d\s}(Q)\times\no
&V_{\m\a\b}(P,-K,K-P)V_{\n\g\d}(-P,K,P-K)V_{\r\h\t}(K'-P',-K',P')V_{\s\l\k}(P'-K',K',-P').
\end{align}
As in the case of Fig.~\ref{g_Coul_prop}, the symmetry factor $1/2$ is compensated by the factor of $2$ from flipping directions of $K'$ and $P'$.
The phase space is defined as $P.S.=(2\p)^4\th(p_0)\th(p_0')\th(k_0)\th(k_0')f_g(p_0)(1+f_q(k_0))f_g(p_0')(1+f_g(k_0'))$. Using \eqref{spin_sum}, we can rewrite the above as
\begin{align}
&\Tr[\P^{\m\n>}(P)D_{\n\m}^<(P)]=\int_{P.S.}\sum_{i,j,k,l}\e^{\a,j}(K)\e^{\g,j*}(K)\e^{\n,i}(P)\e^{\n,i}(P)\e^{\m,i*}(P)\e^{\k,k*}(P')\e^{\t,k}(P')\e^{\l,l*}(K')\e^{\h,l}(K')\times\no
&D_{22}^{\b\r}(Q)D_{11}^{\d\s}(Q)V_{\m\a\b}(P,-K,K-P)V_{\n\g\d}(-P,K,P-K)V_{\r\h\t}(K'-P',-K',P')V_{\s\l\k}(P'-K',K',-P')\no
&=\int_{P.S.}\sum_{\text{spin}}|{\cal T}(P,P',K,K')|^2,
\end{align}
with ${\cal T}(P,P',K,K')$ being t-channel amplitude for Coulomb scattering. Thus it corresponds to $|t|^2$ of Coulomb scattering. The right diagram of Fig.~\ref{g_Coul_prop} is evaluated similarly to give $\frac{1}{2}|s|^2$. In this case, the symmetry factor is not compensated because flipping the directions of $K$ and $K'$ is kinematically not allowed. By relabeling the momenta $K\leftrightarrow K'$ in the first diagram, we can easily show $|s|^2=|u|^2$. Thus the sum gives $|s|^2+\frac{1}{2}|t|^2=\frac{1}{2}(|s|^2+|t|^2+|u|^2)$.

Next we consider the vertex correction diagrams. The first diagram in Fig.~\ref{g_Coul_vert} gives the following contribution
\begin{align}
&\Tr[\P^{\m\n>}(P)D_{\n\m}^<(P)]=\int_{P.S.}D_{22}^{\a\h}(P-K)D_{11}^{\k\d}(P-K')P_T^{\b\l}(K)P_T^{\r\s}(P')P_T^{\t\g}(K')P_T^{\n\m}(P)\times\no
&V_{\m\a\b}(P,K-P,-K)V_{\n\g\d}(-P,K',P-K')V_{\r\h\t}(P',P-K,-K')V_{\s\l\k}(-P',K,K'-P)\no
&=\int_{P.S.}\sum_{i,j,k,l}D_{22}^{\a\h}(P-K)D_{11}^{\k\d}(P-K')V_{\m\a\b}(P,K-P,-K)V_{\n\g\d}(-P,K',P-K')\times\no
&V_{\r\h\t}(P',P-K,-K')V_{\s\l\k}(-P',K,K'-P)\e_i^{\l*}(K)\e_i^\b(K)\e_j^{\g*}(K')\e_j^\t(K')\e_k^\n(P)\e_k^{\m*}(P)\e_l^\s(P')\e_l^{\r*}(P')\no
&=\int_{P.S.}\sum_{\text{spin,color}}{\cal T}(P,P',K,K')^*{\cal U}(P,P',K,K'),
\end{align}
which is the interference term $t^*u$. Similarly, the remaining diagrams give the following interference terms $s^*u$ and $st^*$ respectively. Again by relabeling of momenta $K\leftrightarrow K'$, we can show $t^*u=tu^*$, $s^*u=s^*t$, $st^*=su^*$. Collecting all contributions, we have
\begin{align}
\Tr[\P^{\m\n>}(P)D_{\n\m}^<(P)]=\frac{1}{2}\int_{P.S.}\sum_{\text{spin,color}}|{\cal S}(P,P',K,K')+{\cal T}(P,P',K,K')+{\cal U}(P,P',K,K')|^2.
\end{align}
The symmetry factor matches with that in kinetic theory.
\begin{figure}
	\includegraphics[width=0.45\textwidth]{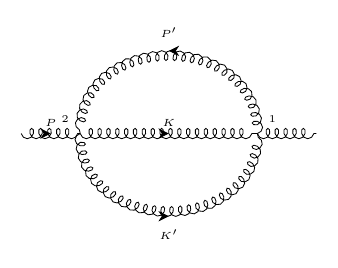}
	\caption{\label{g_contact}Gluon self-energy diagrams with 4-gluon vertices only.}
\end{figure}
\begin{figure}
	\includegraphics[width=0.45\textwidth]{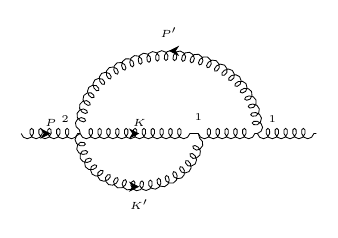}
	\includegraphics[width=0.45\textwidth]{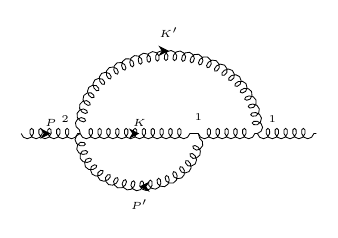}
	\includegraphics[width=0.45\textwidth]{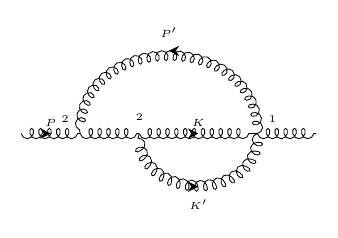}
	\includegraphics[width=0.45\textwidth]{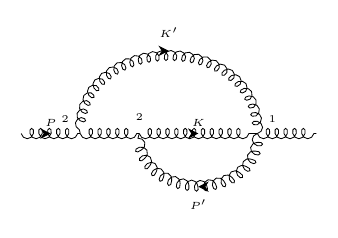}
	\caption{\label{g_mix}Gluon self-energy diagrams with both 3-gluon and 4-gluon vertices.}
\end{figure}
We have left out diagram involving 4-gluon vertices, which can also interfere with the diagrams considered above. The diagrams with only 4-gluon vertices and mixed vertices are shown in Fig.~\ref{g_contact} and Fig.~\ref{g_mix} respectively. The contribution of Fig.~\ref{g_contact} is given by
\begin{align}
&\Tr[\P^{\m\n>}(P)D_{\n\m}^<(P)]=\frac{1}{2}\int_{P.S.}P^{\a\g}_T(P')P^{\n\m}_T(P)P^{\b\d}_T(K)P^{\r\s}_T(K')V_{\m\a\b\r}(P,P',-K,-K')V_{\n\b\d\s}(-P,-P',K,K')\no
&=\frac{1}{2}\int_{P.S.}\sum_{i,j,k,l}\e_i^{\a*}(P')\e_i^\g(P')\e_j^{\m*}(P)\e_j^\n(P)\e_k^{\d*}(K)\e_k^{\b}(K)\e_l^{\s*}(K')\e_l^\r(K')V_{\m\a\b\r}(P,P',-K,-K')V_{\n\b\d\s}(-P,-P',K,K')\no
&=\frac{1}{2}\sum_{\text{spin,color}}|{\cal F}(P,P',K,K')|^2,
\end{align}
where $V_{\m\a\b\r}$ and $V_{\n\b\d\s}$ are 4-gluon vertices, again with color indices suppressed. ${\cal F}(P,P',K,K')$ are scattering amplitude with 4-gluon vertex. We shall refer to this as f-channel, with its amplitude square abbreviated $|f|^2$. The diagrams in Fig.~\ref{g_mix} can be analyzed similarly to give $\frac{1}{2}sf^*$, $uf^*$ $\frac{1}{2}s^*f$ and $u^*f$. Furthermore, the relabeling $K\leftrightarrow K'$ in the second and fourth diagrams of Fig.~\ref{g_mix} gives the relations $uf^*=tf^*$ and $u^*f=t^*f$. Finally, we can combine all square and interference terms to obtain
\begin{align}
\Tr[\P^{\m\n>}(P)D_{\n\m}^<(P)]=\frac{1}{2}\int_{P.S.}\sum_{\text{spin,color}}|{\cal S}(P,P',K,K')+{\cal T}(P,P',K,K')+{\cal U}(P,P',K,K')+{\cal F}(P,P',K,K')|^2.
\end{align}
We already have the loss term from $\Tr[\P^{\m\n>}(P)D_{\n\m}^<(P)]$ for Compton and Coulomb scattering for gluons. Again the gain term is given by $\Tr[\P^{\m\n>}(P)D_{\n\m}^<(P)]$, which only exchanges the initial state with the final state.

\subsection{Inelastic collision}

\subsubsection{Gluon inelastic collision: structure of diagrams}\label{sec:diagram}
The inelastic collisions arise from a subset of multi-loop self-energy diagrams that are enhanced by pinching mechanism. The diagrammatic analysis for QED case has been performed in \cite{Arnold:2001ba} and physical picture for QCD case has been elaborated in \cite{Arnold:2002ja,Arnold:2008iy}. We shall illustrate the structure in QCD case with a two-loop example, and deduce from it the general structure of self-energy beyond two-loop. 
For pedagogical reasons, we first discuss pure glue case, for which a permutation symmetry is present, and then generalize to QCD.
\begin{figure}
	\includegraphics[width=0.45\textwidth]{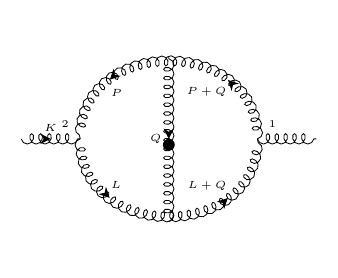}
	\includegraphics[width=0.45\textwidth]{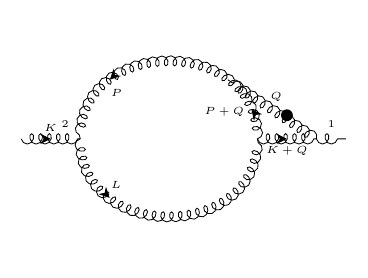}
	\includegraphics[width=0.45\textwidth]{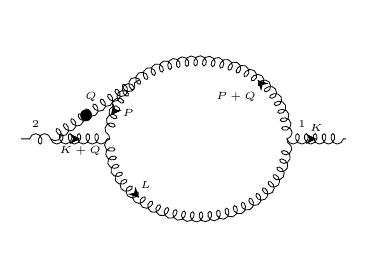}
	\includegraphics[width=0.45\textwidth]{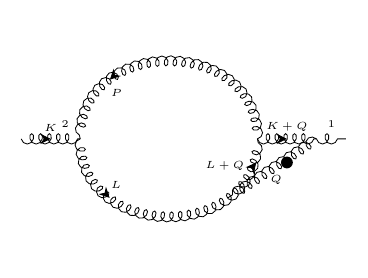}
	\includegraphics[width=0.45\textwidth]{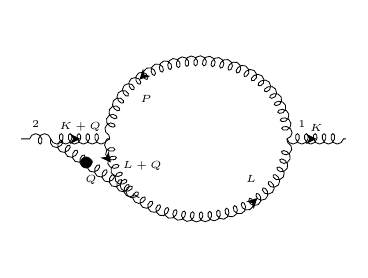}
	\caption{\label{diag}Two loop gluon self-energy diagrams contributing to inelastic collisions. Each diagram has four possible contour labelings. The propagators with heavy dot correspond to resummed ones for soft momentum.}
\end{figure}

\begin{figure}
	\includegraphics[width=0.45\textwidth]{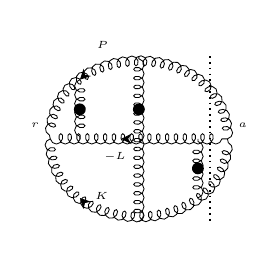}
	\caption{\label{bound}Schematic representation of \eqref{12-ra-2loop}. The $r$ and $a$ labels come from the factor $D_{ra}(P)D_{ra}(-L)D_{ra}(K)$. Note that this is not a conventional diagram in $ra$-basis, in which such labeling is not allowed.}
\end{figure}

The gluon self-energy diagram at two-loop corresponding to inelastic collision is depicted in Fig.~\ref{diag}. It distinguishes from the diagram for elastic collision in the existence of soft propagators. The suppression in phase space can be compensated by pinching mechanism. The first diagram in Fig.~\ref{diag} has the same topology as the photon self-energy in QED case. The other four diagrams are specific to QCD. We aim at deriving a simple representation for the sum of the diagrams. We will use the following relations to switch from the Keldysh basis to the $ra$-basis, in which the pinching integration is most conveniently treated
\begin{align}\label{12-ra}
&D^<(X,P)=f(X,P)\(D_{ra}(X,P)-D_{ar}(X,P)\),\no
&D^>(X,P)=(1+f(X,P))\(D_{ra}(X,P)-D_{ar}(X,P)\),\no
&D^{11}(X,P)=(1+f(X,P))D_{ra}(X,P)-f(X,P)D_{ar}(X,P),\no
&D^{22}(X,P)=f(X,P)D_{ra}(X,P)-(1+f(X,P))D_{ar}(X,P).
\end{align}
These are the equilibrium KMS relation promoted to local distribution $f(X,P)$. The Lorentz indices have been suppressed, as they are inessential for our discussion below. We write down the representation of $\P^>(K)$ for the five diagrams below. Each diagram has four contributions corresponding to four possible labelings of contour indices. The first diagram gives the following integrand (with subscript corresponding to the diagram number)
\begin{align}\label{2loop}
&\P^>_1(K)=D^{11}(P+Q)D^<(P)D^>(L)D^{11}(L+Q)D^{11}(Q)-D^{11}(P+Q)D^<(P)D^{22}(L)D^>(L+Q)D^<(Q)\no
&-D^<(P+Q)D^{22}(P)D^>(L)D^{11}(L+Q)D^>(Q)+D^<(P+Q)D^{22}(P)D^{22}(L)D^>(L+Q)D^{22}(Q)\no
&\simeq\(D^{11}(P+Q)D^<(P)-D^<(P+Q)D^{22}(P)\)\(D^>(L)D^{11}(L+Q)-D^{22}(L)D^>(L+Q)\)D_{rr}(Q),
\end{align}
with $L=K+P$.
We have also dropped explicit vertices, which is fixed by hard momenta in the diagram thus is independent of the location the soft propagators are attached to. We have also suppressed the coordinate $X$ as all the distributions are taken within the same fluid element. The minus sign comes from the relative sign between vertices of type $2$ and type $1$. To the order of our interest, we may approximate $f(Q)\sim T/q_0\gg 1$ for the soft momentum $Q$, thus we may drop the $1$ next to $f$ in \eqref{12-ra} to have $D^<(Q)=D^>(Q)=D^{11}(Q)=D^{22}(Q)=D_{rr}(Q)$. Using \eqref{12-ra}, we can rewrite the above by keeping only the pinching enhanced terms to have \footnote{Only products of all $D_{ra}(D_{ar})$ propagators are pinching enhanced.}
\begin{align}
\P_1^>(K)\sim f(P)f(-L)\(D_{ra}(P+Q)D_{ra}(P)D_{ra}(-L)D_{ra}(-L-Q)+ra\leftrightarrow ar\).
\end{align}
In arriving at the above, we have also used $f(P+Q)\simeq f(P)$ and $f(-L-Q)\simeq f(-L)$.
Multiplying by $D^<(K)$ and using \eqref{12-ra}, we have
\begin{align}
&\P_1^>(K)D^<(K)\sim \(D_{ra}(K)-D_{ar}(K)\)\big[D_{ra}(P+Q)D_{ra}(P)D_{ra}(-L)D_{ra}(-L-Q)+ra\leftrightarrow ar\big]\no
&f(P)f(-L)f(K).
\end{align}
The appearance of distribution functions suggests a kinetic interpretation with $K$ and $P$ being incoming momenta and $L$ being outgoing momentum (assuming $k_0,\,p_0>0$ to be specific). The square bracket corresponds to soft momentum exchange beween $P$ and $-L$, which will be interpreted as interaction between constituents with momentum $P$ and $-L$. It is natural to expect interaction involving constituent with momentum $K$ as suggested by the diagrams. We illustrate with the second and third diagrams, which give
\begin{align}
&\P_2^>(K)\sim\big[\(D^{11}(P-Q)D^<(P)-D^<(P-Q)D^{22}(P)\)D^>(L)D^{11}(K+Q)\no
&+\(-D^>(P-Q)D^<(P)+D^{22}(P-Q)D^{22}(P)\)D^{22}(L)D^>(K+Q)\big]D_{rr}(Q),\no
&\P_3^>(K)\sim\big[\(D^{11}(P-Q)D^{11}(P)-D^<(P-Q)D^>(P)\)D^{11}(L)D^>(K+Q)\no
&+\(-D^{11}(P-Q)D^<(P)+D^<(P-Q)D^{22}(P)\)D^>(L)D^{22}(K+Q)\big]D_{rr}(Q).
\end{align}
Terms proportional to $D^{11}(P-Q)D^{11}(P)$ and $D^{22}(P-Q)D^{22}(P)$ have no obvious kinetic interpretation. In fact, they cancel when we take the sum in the collision term $\P^>(K)D^<(K)-\P^<(K)D^>(K)$. To see that, we note $\P^<(K)D^>(K)$ can be obtained from $\P^>(K)D^<(K)$ by flipping the indices $1\leftrightarrow 2$ and $>\leftrightarrow<$. This allows us to drop these terms to have
\begin{align}
&(\P^>_2(K)+\P^>_3(K))D^<(K)\sim \(D_{ra}(K)-D_{ar}(K)\)\big[D_{ra}(P-Q)D_{ra}(P)D_{ra}(-L)D_{ra}(K+Q)\no
&+ra\leftrightarrow ar\big]f(P)f(-L)f(K).
\end{align}
Similarly we have from the last two diagrams
\begin{align}
&(\P^>_4(K)+\P^>_5(K))D^<(K)\sim \(D_{ra}(K)-D_{ar}(K)\)\big[D_{ra}(-L)D_{ra}(-L-Q)D_{ra}(P)D_{ra}(K+Q)\no
&+ra\leftrightarrow ar\big]f(P)f(-L)f(K).
\end{align}
Taking the sum of all, and keeping the pinching enhanced terms, we can rewrite the result in a more symmetric form as
\begin{align}\label{12-ra-2loop}
&\P^>(K)D^<(K)\sim f(P)f(-L)f(K)\big[D_{ra}(P)D_{ra}(-L)D_{ra}(K)\big(D_{ra}(P+Q)D_{ra}(-L-Q)+\no
&D_{ra}(P_Q)D_{ra}(K+Q)+D_{ra}(-L-Q)D_{ra}(K+Q)\big)-ra\leftrightarrow ar\big]\no
&=-f(P)(1+f(L))f(K)2\text{Re}\big[D_{ra}(P)D_{ra}(-L)D_{ra}(K)\big(D_{ra}(P+Q)D_{ra}(-L-Q)+\no
&D_{ra}(P-Q)D_{ra}(K+Q)+D_{ra}(-L-Q)D_{ra}(K+Q)\big)\big],
\end{align}
where we have used $D_{ar}=-D_{ra}^*$. Now it is manifestly symmetric with respect to the three momenta, with the three terms in the bracket corresponding to pairwise interactions. Note that despite the external momentum $K$ is fixed. The soft interaction doesn't change $K$ at $O(1)$.
We represent \eqref{12-ra-2loop} diagrammatically in Fig.~\ref{bound}. The two loop example leads to a first order perturbation of pairwise interaction in a three-body system. It naturally generalizes to multiple interactions, which can be resummed into an integral equation, as we will derive below.

\subsubsection{Resummed 3-gluon vertex}\label{sec:3g}

To set the stage, we first note the diagrams with arbitrary insertions of soft gluon exchange can be split into a part with one resummed and the other part with one bare vertex as shown by the dotted line in Fig.~\ref{bound}. The three momenta are collinear up to $O(g)$, which can be decomposed into a longitudinal and transverse components. To make the decomposition unbiased, we choose a longitudinal unit vector $\he$, which is collinear with the three momenta up to $O(g)$, but is otherwise arbitrary. We will derive an integral equation for the resummed vertex below. Unlike in the previous section, we will not perform spin sum, but use a spin basis in the derivation. The spin basis will allow us to discuss conversion between spin and orbital angular momentum (OAM) with $\he$ naturally chosen as the quantization axis.

The resummed vertex involves contracting the bare vertex with three on-shell gluon propagators. The transverse property of the on-shell gluon forces the resummed vertex to carry transverse indices only. The following contraction will be needed
\begin{align}
P_{im}^T(P)P_{jn}^T(L)P_{kl}^T(K)\[\d_{ik}(-p+k)_j+\d_{kj}(-k-l)_i+\d_{ji}(l+p)_k\],
\end{align}
with the projectors from propagators and square bracket from the bare vertex. The color structure will be considered later. We find it more instructive to split the projectors into halves by the following relation
\begin{align}\label{pol_sum}
P_{im}^T=\sum_s\e_i^s\e_m^{s*}=\sum_s\e_i^{-s}\e_m^{-s*},
\end{align}
and assign the polarization vectors pair for contraction with both resummed and bare vertices respectively, so that both vertices are forced to be transverse. We use the following explicit expression for polarization vectors
\begin{align}\label{pol_vec}
\e^s(L)=\sqrt{2}\(1,is,-\frac{l^s}{l_\pr}\),
\end{align}
with $s=\pm1$ corresponding helicity of the gluon. $l^s\equiv l_1+isl_2$ and $l_\pr={\vec l}\cdot\he$. The tiny longitudinal component is needed to account for the deviation of ${\hat l}$ from $\he$. Note that $\e^s$ is normalized up to $O(g)$ because $l_\pp\sim O(g)$. We calculate the following contraction
\begin{align}\label{vert_contr}
&\e_i^s(P)\e_j^{-s'*}(L)\e_k^{s''}(K)\[\d_{ik}(-p+k)_j+\d_{kj}(-k-l)_i+\d_{ji}(l+p)_k\]\no
&=\d_{s,-s''}\frac{1}{\sqrt{2}}\[(-p+k)^{s'}-\frac{-\ppr+\kpr}{\lpr}l^{s'}\]+\d_{-s',s''}\frac{1}{\sqrt{2}}\[(-k+l)^{s}-\frac{-\kpr-\lpr}{\ppr}p^{s}\]\no
&+\d_{s,-s'}\frac{1}{\sqrt{2}}\[(l+p)^{s''}-\frac{\lpr+\ppr}{\kpr}k^{s''}\]\no
&=\d_{s,-s''}{\sqrt{2}}\[\frac{\ppr k_\pp-\kpr p_\pp}{\lpr}\]^{s'}+\d_{-s',s''}\sqrt{2}\[\frac{\kpr l_\pp-\lpr k_\pp}{\ppr}\]^{s}+\d_{s,-s'}\sqrt{2}\[\frac{\lpr p_\pp-\ppr l_\pp}{\kpr}\]^{s''}.
\end{align}
With our specific kinematics $p_0,\,k_0>0$, the momenta $P$ and $K$ are incoming and $L$ is outgoing. We have chosen helicity $s$ and $s''$ for the initial state and helicity $-s'$ for the final state. This is consistent with crossing symmetry and leads to the following permutation symmetry: $p\to k\to-l\to p$ and $s\to s''\to s'\to s$, which relates the three terms in \eqref{vert_contr}.
The contraction \eqref{vert_contr} is naturally interpreted as amplitude of an elementary gluon fusion process. 

We pause to discuss the conservation of angular momentum. Clearly the gluon fusion process don't conserve spin, so naively the conservation of angular momentum is violated. But this process also involves change of OAM as encoded in the amplitude. Each of three terms in \eqref{vert_contr} corresponds to a different pattern of spin-orbit conversion, which we elaborate now. It is convenient to discuss the conservation of angular momentum along $\he$. To $O(g^0)$, we don't distinguish the direction of the collinear momenta. It is easy to deduce the change of spin projection along $\he$ corresponding to the first term of \eqref{vert_contr} to be $\D S=-s'$. The combination $\frac{\ppr k_\pp-\kpr p_\pp}{\lpr}$ can be interpreted as the relative momentum between the two gluons in the initial state, which becomes exact if we align $\he$ with $\hat{l}$ so that $k_\pp=-p_\pp$. We can then read out the OAM projection by look at the dependence of the amplitude on the relative momentum. If the initial state is an OAM eigenstate $|l,m\>$, the dependence would be $\<l,m|\hat{q}\>\sim Y_{l,m}^*(\hat{q})$ with $q$ denoting the relative momentum. The specific amplitude is proportional to $q_1+is q_2\sim Y_{1,-s}^*$ indicating $\D L=s'$. Thus conservation of angular momentum indeed holds. The other two terms can be simply obtained by permutation symmetry to give $\D S=-s,\,\D L=s$ and $\D S=-s'',\,\D L=s''$ respectively.

\begin{figure}
	\includegraphics[width=0.3\textwidth]{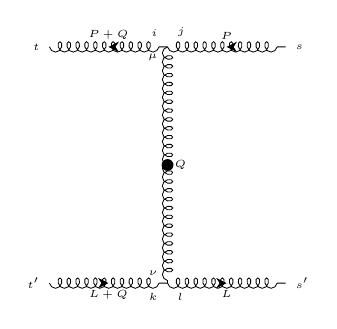}
	\caption{\label{ladder}Soft gluon propagator contracted with four polarization vectors each from ``half'' of the projector of hard on-shell propagator, see text for details.}
\end{figure}

Now let's consider the effect of inserting one soft gluon exchange, which we choose between $P$ and $-L$ to be specific. Since we have reserved half of the projector to the vertex, we should consider contraction of the soft gluon exchange with the other half of the projector on two side as shown in Fig.~\ref{ladder}. The explicit contraction without color factors is as follows
\begin{align}\label{exch_contr}
&g^2\e_i^{t*}(P+Q)e_j^{s}(P)\[\h^{i\m}(-p-2q)^j+\h^{\m j}(q-p)^i+\h^{ij}(2p+q)^\m\]D_{rr,\m\n}^*(Q)\e_k^{-t'}(L+Q)\e_l^{*-s'}\no
&\[\h^{k\n}(l+2q)^l+\h^{\n l}(-q+l)^k+\h^{lk}(-2l-q)^\n\]\no
&\simeq -2g^2P^\m\d_{st}D_{rr,\m\n}^*(Q)2L^\n\d_{s't'}
\simeq -4g^2p_\pr l_\pr e^\m e^\n D_{rr,\m\n}^*(Q)\d_{st}\d_{s't'}.
\end{align}
We see that exchange of soft momentum between hard on-shell particles don't flip their spin. It follows that the spin structure of the vertex is preserved. The color factor is calculated separately as (referring to Fig.~\ref{color} for color indices.)
\begin{align}\label{color_contr}
\sum_{d,a,b}f_{adm}f_{bdn}f_{abc}=\frac{1}{2}C_Af_{mnc}.
\end{align}
It shows that the color structure is also preserved.
\begin{figure}
	\includegraphics[width=0.45\textwidth]{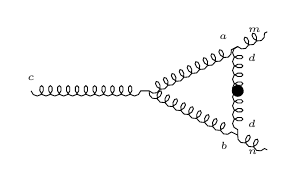}
	\caption{\label{color}Representation of color indices in one insertion of soft gluon propagator.}
\end{figure}

\begin{figure}
	\includegraphics[width=0.45\textwidth]{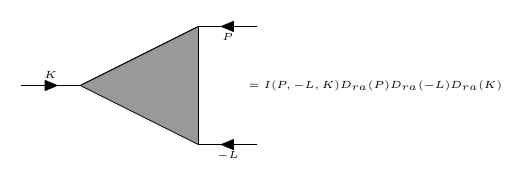}
	\\[1em]
	\includegraphics[width=0.7\textwidth]{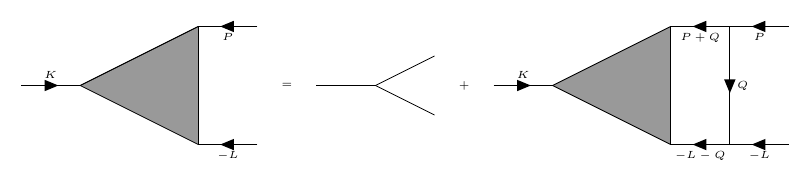}
	\caption{\label{color}Graphic representation of resummed vertex and resummation equation with soft gluon exchange between one particular pair.}
\end{figure}
We are ready to derive an integral equation for the resummed vertex. Since spin structure is preserved, we shall suppress the spin indices. In view of \eqref{12-ra-2loop}, we parametrize the retarded component of the resummed vertex as $I(P,-L,K)D_{ra}(P)D_{ra}(-L)D_{ra}(K)$. $D_{ra}(P)$, $D_{ra}(-L)$ and $D_{ra}(K)$ are scalar retarded propagators extracted from the gluons stretching from the vertex. Half of the projectors in the gluon propagators have been contracted as in \eqref{vert_contr} to give $I(P,-L,K)$. The graphic representation of resummation equation with soft gluon exchange between $P$ and $-L$ is given explicitly as
\begin{align}
&I(P,-L,K)D_{ra}(P)D_{ra}(-L)D_{ra}(K)=I_0(P,-L,K)D_{ra}(P)D_{ra}(-L)D_{ra}(K)+g^2\frac{C_A}{2}\times\no
&\int_Q I(P+Q,-L_Q,K)D_{ra}(P+Q)D_{ra}(-L-Q)D_{ra}(K)D_{ra}(P)D_{ra}(-L)(-4p_\pr l_\pr) \he^\m \he^\n D_{rr,\m\n}^*(Q),
\end{align}
with the additional factor from \eqref{exch_contr} and \eqref{color_contr}.
To proceed, we apply $\int_{p_0,k_0}\equiv\int\frac{dp_0dk_0}{(2\p)^2}$ to pick up the pinching pole contributions. Unlike the QED case, the poles of all three momenta are shifted from the real axis. Performing the contour integration carefully in appendix, we obtain
\begin{align}\label{F_eq}
&\text{LHS}=\frac{I(P,-L,K)}{-8\ppr\lpr\kpr(i\d E+3\G/2)}\equiv \frac{F(P,-L,K)}{-8\ppr\lpr\kpr},\no
&\text{RHS}=\frac{1}{-8\ppr\lpr\kpr(i\d E+3\G/2)}I_0(P,-L,K)+\frac{1}{-8\ppr\lpr\kpr}\int_Q F(P+Q,-L-Q,K)\frac{1}{-4\ppr\lpr(i\d E+3\G/2)}\no
&g^2\frac{C_A}{2}\(-4\ppr\lpr D_{\m\n}(Q)\he^\r\he^\n\),
\end{align}
with $\d E=\frac{p_\pp^2+m_g^2}{2\ppr}+\frac{l_\pp^2+m_g^2}{-2\lpr}+\frac{k_\pp^2+m_g^2}{2\kpr}$.
Multiplying $i\d E+3\G/2$ to both sides, and adding the other two contributions from soft gluon exchange between other two momenta, we obtain
\begin{align}
&(i\d E+\frac{3\G}{2})F(P,-L,K)=I_0(P,-L,K)+g^2\frac{C_A}{2}\int_Q 2\p\d(q_0-q_\pr) D_{rr,\m\n}^*(Q)\he^\m\he^\n\big[F(P+Q,-L-Q,K)+\no
&F(P+Q,-L,K-Q)+F(P,-L-Q,K+Q)\big].
\end{align}
The factor of $2\p\d(q_0-q_\pr)$ is introduced because the integrations of $\int_{p_0}$ and $\int_{p_0+q_0}$ have localized $q_0$ at $p_\pr+q_\pr-p_\pr=q_\pr$.
Using \eqref{Gamma_g}, we can rewrite \eqref{F_eq} in the following form
\begin{align}\label{3-body}
&i\d E F(P,-L,K)=I_0(P,-L,K)+g^2\frac{C_A}{2}\int_Q 2\p\d(q_0-q_\pr) D_{rr,\m\n}^*(Q)\he^\m\he^\n\Big(\[F(P+Q,-L-Q,K)-F(P,-L,K)\]+\no
&\[F(P+Q,-L,K-Q)-F(P,-L,K)\]+\[F(P,-L-Q,K+Q)-F(P,-L,K)\]\Big).
\end{align}
As is already noted in \cite{Arnold:2008iy}, this equation has the interpretation of 2D Schr\"odinger equation for a 3-body system if we ignore the inhomogeneous term for now. The masses and momenta of three constituents are given by $(\ppr,-\lpr,\kpr)$ and $(p_\pp,-l_\pp,k_\pp)$ respectively. A unique feature of the system is that the total mass and momentum is zero. A consequence of this is that there is only one independent relative momentum, which can be chosen as any one of the following appearing in \eqref{vert_contr} already
\begin{align}
{\bf p}_{12}=\frac{\kpr{\bf \ppp}-\ppr{\bf \kpp}}{\lpr},\quad {\bf p}_{23}=\frac{\lpr{\bf \kpp}-\kpr{\bf \lpp}}{\ppr},\quad {\bf p}_{13}=\frac{\ppr{\bf \lpp}-\lpr{\bf \ppp}}{\kpr}.
\end{align}
In particular, ${\bf p}_{12}$ is related to the ${\bf h}$ vector defined in \cite{Arnold:2008iy} as ${\bf p}_{12}=\frac{{\bf h}}{\lpr}$. The other two momenta are related as ${\bf p}_{13}=-\frac{\lpr}{\kpr}{\bf p}_{12}$ and ${\bf p}_{23}=-\frac{\lpr}{\ppr}{\bf p}_{12}$. $\d E$ has the interpretation of non-relativistic kinetic energy if we ignore $m_g$. The pairwise gluon exchange induced interaction is purely dissipative as it has a relative $i$ with respect to the kinetic energy. The dissipative nature of the interaction is crucial for nonvanishing vertex: in the absence of interaction and with real $I_0$, $F(P,-L,K)$ would be purely imaginary. Since the complete vertex is obtained by taking the real part of $F$ from \eqref{12-ra-2loop}, we would have vanishing vertex without interaction. Taking the real part gives the spectral density of the 3-body system, which vanishes in the absence of dissipation.

We can actually further reduce the problem given that there is only one independent kinematic variable. Let's further write
\begin{align}
I_0(P,-L,K)=-\d_{s,-s''}\sqrt{2}\frac{h^{s'}}{\lpr}+\text{perm}.
\end{align}
This motivates the following parametrization
\begin{align}
F_{s,-s',s''}(P,-L,K)=-\d_{s,-s''}\sqrt{2}\frac{V^{s'}}{\lpr}+\text{perm},
\end{align}
in terms of a 2D vector $V^{s'}(\bh)$ resulting from interaction effect.
We have restored the spin configuration. By permutation symmetry on \eqref{3-body}, it is not difficult to deduce the following equation should hold
\begin{align}\label{V_eq}
&-i\d E V^{s'}(\bh)=-h^{s'}-g^2\frac{C_A}{2}\int_Q D_{rr,\m\n}^*(Q)\he^\m\he^\n\Big(V^{s'}(\bh+(1-x){\bf q}_\perp)+V^{s'}(\bh+x {\bf q}_\pp)\no
&+V^{s'}(\bh+{\bf q}_\pp)-3V^{s'}(\bh)\Big),
\end{align}
with $x=\kpr/\lpr$. Note that $D_{rr,\m\n}^*\he^\m\he^\n$ is unpolarized. The equations for $s'=\pm1$ are degenerate and equivalent to a 2D vector equation, which is equivalent to the one written down in \cite{Arnold:2008iy}, with the identification ${\bf V}=\frac{1}{2}{\bf F}_{\text{AMY}}$. The factor of $1/2$ comes from the spin average.

So far we have restricted to fusion kinematics with $p_0,\,k_0>0$. The situation of other kinematics can be easily deduced: by the property $\e^s(P)=\e^{-s*}(-P)$, we can simply move a particle with negative momentum from initial to final state (or vice versa) with its helicity flipped. Since we have chosen the initial/final particle to carry helicity $\pm s$, the equations in the above are equally applicable for other kinematics.

Finally we can write down the loss component of the collision term. restoring color and symmetry factors in \eqref{12-ra-2loop} and converting the Lorentz indices contraction into spin sum, we have
\begin{align}\label{loss_gluon_loop}
&\P^{\m\n>}(K)D_{\n\m}^<(K)=\frac{1}{2}g^2C_A\int_P f(P)f(K)f(-L)2\text{Re}\sum_{s,s's''}\Big[D_{ra}(P)D_{ra}(-L)D_{ra}(K)I^{s,-s',s''}(P,-L,K)\no
&\times I_0^{s,-s',s''*}(P,-L,K)\Big].
\end{align}
The complex conjugate on the bare vertex follows from the way we split the projectors, in which we assign amplitude to resummed vertex and the conjugate amplitude to the bare vertex.
In order to convert \eqref{Boltzmann_g} into the Boltzmann form, we need to perform the energy integration. On the LHS, the shift of dispersion is a subleading effect. On the RHS, we use the following integration
\begin{align}\label{IF}
\int_{p_0,k_0}D_{ra}(P)D_{ra}(-L)D_{ra}(K)I(P,-L,K)\simeq\frac{F(P,-L,K)}{-8\ppr\lpr\kpr}.
\end{align}
The spin sum can be performed using
\begin{align}
&\sum_{s,s's''}F^{s,-s',s''}(P,-L,K)I_0^{s,-s',s''*}(P,-L,K)\no
&=\sum_{s,s's''}2\(\d_{s,-s''}V^{s'}-\d_{-s',s''}V^s\frac{1}{x}-\d_{s,-s'}V^{s''}\frac{1}{1-x}\)\(\d_{s,-s''}h^{-s'}-\d_{-s',s''}h^{-s}\frac{1}{x}-\d_{s,-s'}h^{-s''}\frac{1}{1-x}\)\no
&=8{\bf V}\cdot \bh\(1+\frac{1}{x^2}+\frac{1}{(1-x)^2}\).
\end{align}
Although there is apparent cross terms between different spin configurations, the cross terms cancel out to give a sum of three square amplitudes.
The final result is given by
\begin{align}
&\int_{P,k_0}\P^{\m\n>}(K)D_{\n\m}^<(K)=g^2C_A\int_{\ppr,\ppp} f(P)f(K)f(-L)\frac{1}{-8\ppr\lpr\kpr}\text{Re}[8{\bf V}\cdot \bh]\(1+\frac{1}{x^2}+\frac{1}{(1-x)^2}\)\no
&=g^2C_A\int_{\ppr,h} f(\ppr)f(\kpr)(1+f(\lpr))\frac{1}{\ppr\lpr \kpr^3}\text{Re}[{\bf V}\cdot \bh]\(1+\frac{1}{x^2}+\frac{1}{(1-x)^2}\).
\end{align}
We have switched the integration variable from $\ppp$ to $h$, and replaced the argument in the distribution function by longitudinal momenta.

\subsubsection{Resummed quark-gluon vertex}

We then turn to quark loop diagram. It is useful to revisit the QED analog in spin basis. Although the interaction in QCD case is more complicated, the spin configurations for quark-gluon vertex and electron-photon vertex are the same. To be specific, we consider fusion kinematics with $p_0,\,k_0>0$. As in the 3-gluon vertex case, we contract the vertex with half of the projectors from the fermion and photon respectively. The spitting of the projectors are defined in \eqref{spin_sum}. We will consider the spin configuration of the contraction $\bar{u}_s(L)\g^i u_t(P)\e^i_r(K)$
We use the following representation for spinor
\begin{align}
u_s(P)=\(\frac{P\cdot\s+m}{\sqrt{2(p_0+m)}}\x_s,\frac{P\cdot\bar{\s}+m}{\sqrt{2(p_0+m)}}\x_s\)^T,
\end{align}
with $\x_s$ being normalized spinor doublet.
Vertices attached to soft photon simply preserve spin as
\begin{align}
\bar{u}_s(P+Q)\g^\m u_{s'}(P)\simeq 2P^\m\d_{ss'}.
\end{align}
Vertices attached to hard on-shell photon can flip spin. We need the following spinor contraction
\begin{align}
\bar{u}_s(L)\g^i u_t(P)\simeq\frac{1}{2\sqrt{l_0p_0}}\[2\x_s^\dg\(\s^i p\cdot\s l_0+l\cdot\s\s^i p_0\)\x_t+2m\x_s^\dg\(\s^i p\cdot\s+l\cdot\s\s^o\)\x_t\].
\end{align}
For the inelastic collision to be relevant, we assume $m\sim O(p_\pp)$ and have dropped $m$ next to $p_0$ and $l_0$. A careful expansion of the above is needed. Using
\begin{align}
\x_s^\dg\s^i p\cdot\s\x_t=\d_{st}\((\he^i\ppr+\ppp^i)+is\ppp^j\e^{ij}_\pp\)-\d_{s,-t}i\sqrt{2}\ppr\e^{ik}_\pp\e_{-s}^k,
\end{align}
with $\e_\pp^{ij}$ being the 2D Levi-Civita symbol and contracting with $\e_r^i(K)$ defined in \eqref{pol_vec}, we obtain the following nonvanishing spin configurations
\begin{align}\label{spin_qg}
&\bar{u}_s(L)\g^i u_s(P)\e_s^i=\(\frac{2\lpr}{\ppr}\)^{1/2}\frac{\kpr p_s-\ppr k_s}{\kpr},\quad \D S=-s,\,\D L=s\no
&\bar{u}_s(L)\g^i u_s(P)\e_{-s}^i=\(\frac{2\lpr}{\ppr}\)^{1/2}\frac{\kpr p_{-s}-\ppr k_{-s}}{\kpr},\quad \D S=s,\,\D L=-s\no
&\bar{u}_s(L)\g^i u_{-s}(P)\e_s^i=\(\frac{1}{2\lpr\ppr}\)^{1/2}2m\kpr.\quad \D S=\D L=0.
\end{align}
The last one is missed in the previous study \cite{Lin:2021mvw}.
As in the gluon fusion case, the spins of initial/final particles are projected along $\he$. The OAM is deduced from the dependence on relative momentum in the initial state. With $\<l,m|\hat{p}\>\propto Y_{l,m}^*(\hat{p})$ and the property of spherical harmonics $Y_{1,s}(\hat{p})\propto p_s$ and $Y_{1,0}(\hat{p})\propto \ppr$, we easily deduc the OAM in the initial state. The changes of spin and OAM are marked in \eqref{spin_qg}. We see the conservation of total angular momentum. It is worth noting that while the mass term flips spin of the spinor, it actually preserves the total spin. On the other hand, the spin preserving contribution to the vertex induces spin-orbit conversion.

Since the spin configurations are orthogonal to each other, we may derive separate equations for the resummed vertex with specific spin configuration. Let's parametrize the resummed vertices as
\begin{align}\label{Gamma_ABC}
&\bar{u}_s(L)\g^i u_s(P)\e_s^i(K)=\(\frac{2\lpr}{\ppr}\)^{1/2}\frac{\kpr p_s-\ppr k_s}{\kpr}\to\(\frac{2\lpr}{\ppr}\)^{1/2}\frac{\kpr p_s-\ppr k_s}{\kpr}\G_s^A(P,-L), \no
&\bar{u}_s(L)\g^i u_s(P)\e_{-s}^i(K)=\(\frac{2\lpr}{\ppr}\)^{1/2}\frac{\kpr p_{-s}-\ppr k_{-s}}{\kpr}\to\(\frac{2\lpr}{\ppr}\)^{1/2}\frac{\kpr p_{-s}-\ppr k_{-s}}{\kpr}\G_s^B(P,-L), \no
&\bar{u}_s(L)\g^i u_{-s}(P)\e_s^i(K)=\(\frac{1}{2\lpr\ppr}\)^{1/2}2m\kpr\to\(\frac{1}{2\lpr\ppr}\)^{1/2}2m\kpr\G_s^C(P,-L),
\end{align}
with $\G_s^{A,B,C}$ resulting from interaction effect. We could have pointed $\he$ exactly along $\hat{k}$ as photon can't participate in soft photon exchange. For later generalization to QCD, we still choose $\he$ to be almost collinear but otherwise arbitrary. For $\G_s^A$, The soft momentum exchange between $P$ and $-L$ gives rise to the following equation
\begin{align}
&\(\kpr p^s-\ppr k^s\)\G_s^A(P,-L)=\(\kpr p^s-\ppr k^s\)D_{ra}(P)D_{ra}(-L)-e^2\int_Q 2\p\d(q_0-q_\pr) D_{rr,\m\n}^*(Q)\he^\m\he^\n \no
&\times 4\ppr\lpr D_{ra+Q}(P)D_{ra}(-L-Q)\(\kpr (p+q)^s-p k^s\)\G_s^A(P+Q,-L-Q)D_{ra}(P)D_{ra}(-L).
\end{align}
The simplification of the above is similar to the 3-gluon vertex case. We again integrate over $p_0$ and use
\begin{align}\label{qed_int}
\int_{p_0}D_{ra}(P)D_{ra}(-L)=\frac{i}{4\ppr\lpr(\d E'+k_0-\kpr-i\G)}=\frac{1}{4\ppr\lpr(i\d E+\G)}.
\end{align}
The QED counterparts of the parameters are
\begin{align}\label{qed_para}
&\d E'=\frac{p_\pp^2+m_e^2}{2\ppr}+\frac{l_\pp^2+m_e^2}{-2\lpr},\no
&\d E=\frac{p_\pp^2+m_g^2}{2\ppr}+\frac{l_\pp^2+m_g^2}{-2\lpr}+\frac{k_\pp^2+m_\g^2}{2\kpr},\no
&\G=e^2\int_Q D_{rr,\m\n}^*(Q)\he^\m\he^\n.
\end{align}
In the last equality of \eqref{qed_int}, the pole $k_0=\kpr+\frac{\kpp^2+m_\g^2}{2\kpr}$ of $D_{ra}(K)$ has been taken. Note that photon has thermal mass but no thermal width, thus the coefficient in front of $\G$ differs from the QCD counterpart. We denote
\begin{align}
\int_{p_0}\G_s^A=\frac{\c_s}{-4\ppr\lpr},
\end{align}
and express the integrated equation as
\begin{align}
&\(\kpr p^s-\ppr k^s\)\(i\d E+\G\)\c_s(P,-L)=\(\kpr p^s-\ppr k^s\)+e^2\int_Q 2\pi\d(q_0-q_\pr) D_{rr,\m\n}^*(Q)\he^\m\he^\n\(\kpr (p+q)^s-\ppr k^s\)\no
&\times\c_s(P+Q,-L-Q).
\end{align}
Note that $D_{rr,\m\n}^*(Q)\he^\m\he^\n$ is unpolarized, so the equations with $s=\pm1$ are degenerate and equivalent to a 2D vector equation. It may be rewrite in terms of $\bh=\kpr {\bf p}_\perp-\ppr \bkpp$. Using explicit representation of $\G$, we obtain
\begin{align}\label{chi_eq}
h^i i\d E\c(\bh)=h^i+e^2\int_Q 2\pi\d(q_0-q_\pr) D_{rr,\m\n}^*(Q)\he^\m\he^\n\(\(h^i+\kpr q_\pp^i\)\c(\bh+\kpr{\bf q}_\pp)-h^i\c(\bh)\),
\end{align}
where we have expressed the argument of $\c$ in terms of relative momentum $\bh$.
This agrees with \cite{Arnold:2008iy} with the identification $\bppp\c=\frac{1}{2}{\bf f}_\pp^{\text{AMY}}$. The factor $1/2$ is again from spin average. The other half is given by $\G_s^B$, which satisfies exactly the same equation.

$\G_s^C$ satisfies a new equation as
\begin{align}
&2m\kpr s\G_s^C(P,-L)=2m\kpr s D_{ra}(P)D_{ra}(-L)-e^2\int_Q 2\pi\d(q_0-q_\pr) D_{rr,\m\n}^*(Q)\he^\m\he^\n 4\ppr\lpr D_{ra}(P+Q)D_{ra}(-L-Q)\times\no
&2m\kpr s\G_s^C(P+Q,-L-Q)D_{ra}(P)D_{ra}(-L).
\end{align}
Denoting
\begin{align}
\int_{p_0}\G_s^C=\frac{\zeta_s}{-4\ppr\lpr},
\end{align}
and following similar derivation as above, we obtain
\begin{align}\label{zeta_eq}
i\d E\zeta_s(\bh)=1+e^2\int_Q 2\pi\d(q_0-q_\pr) D_{rr,\m\n}^*(Q)\he^\m\he^\n\(\zeta_s(\bh+\kpr{\bf q}_\pp)-\zeta_s(\bh)\).
\end{align}
We have again switched the argument of $\zeta_s$ to relative momentum $\bh$.
It is a scalar equation rather than a vector equation. We may again drop the spin index $s$.

We have restricted to the kinematics with spinors only. The generalization to anti-spinor involves only minimal changes. Let's illustrate with the kinematics $p_0<0,\,l_0>0$, which corresponds to pair production $\g(k)\to\bar{e}(-p)+e(l)$. The contracted vertex in this case is $\bar{u}_s(L)\g^i v_t(-P)\e_r^i(K)$. We use the following representation of anti-spinor
\begin{align}\label{v_rep}
v_t(P)=\(\frac{-P\cdot\s+m}{\sqrt{2(-p_0+m)}}\x_t,\frac{-P\cdot\bar{\s}+m}{\sqrt{2(-p_0+m)}}\x_t\)^T.
\end{align}
It satisfies $v_t(P)=u_t(-P)$ so that $\bar{u}_s(L)\g^i v_t(-P)\e_r^i(K)=\bar{u}_s(L)\g^iu_t(P)\e_r^i(K)$. The expressions in the discussion of spinor case is still applicable, except that the square root in \eqref{v_rep} contains a phase ambiguity. This ambiguity will be canceled in the product of amplitude from resummed vertex and conjugate amplitude from bare vertex.

We are ready to express the loss component of collision term in terms of $\c$ and $\zeta$ with full phase space
\begin{align}
&\P^{\m\n>}(K)D_{\n\m}^<(K)=2e^2f(K)\text{Re}[D_{ra}(K)]\int_P f(P)f(-L)2\text{Re}\sum_{s}\Big[D_{ra}(P)D_{ra}(-L)\no
&\Big(\(\bar{u}_s(L)\g^i u_s(P)\e_s^i(K)\G_s^A\)\(\bar{u}_s(P)\g^j u_s(L)\e_{s}^{j*}(K)\)+\(\bar{u}_s(L)\g^i u_s(P)\e_{-s}^i(K)\G_s^B\)\(\bar{u}_s(P)\g^j u_s(L)\e_{-s}^{j*}(K)\)\no
&+\(\bar{u}_s(L)\g^i u_{-s}(P)\e_s^i(K)\G_s^C\)\(\bar{u}_{-s}(P)\g^j u_{s}(L)\e_s^{j*}(K)\)\Big)
\Big],
\end{align}
with the three terms from square amplitudes of the orthogonal spin configurations. We have expressed $D_{\n\m}^<(K)$ using local KMS relation and the projector $P_{\n\m}^T(K)$ is split into halves as before for contraction in the above \footnote{In this case, $D_{ra}(K)$ doesn't participate in the pinching mechanism, so can be pulled out from the real part as opposed to the QCD case.}. Using \eqref{spin_qg}, we can easily perform the spin sum as
\begin{align}
&\sum_s|\bar{u}_s(L)\g^i u_s(P)\e_s^i(K)|^2=4\frac{\lpr}{\ppr}\frac{h^2}{\kpr^2},\no
&\sum_s|\bar{u}_s(L)\g^i u_s(P)\e_{-s}^i(K)|^2=4\frac{\ppr}{\lpr}\frac{h^2}{\kpr^2},\no
&\sum_s|\bar{u}_s(L)\g^i u_{-s}(P)\e_s^i(K)|^2=\frac{4m^2\kpr^2}{\lpr\ppr}.
\end{align}
We then arrive at after the $p_0$-integration and using definitions of $\c$ and $\zeta$
\begin{align}
&\P^{\m\n>}(K)D_{\n\m}^<(K)=2e^2f(K)\text{Re}[D_{ra}(K)]\int_{\ppr,\bh} f(\ppr)(1+f(\lpr))\frac{1}{\kpr^2}2\text{Re}\Big[
\frac{\lpr^2+\ppr^2}{\lpr^2\ppr^2}\bh^2\c
+m^2\frac{\kpr^2}{\lpr^2\ppr^2}\zeta
\Big],
\end{align}
We have switched the integration measure from $\ppp$ to $h$, which introduces the factor $1/\kpr^2$. The two terms in the square bracket corresponding to spin-preserving and spin-flipping contributions.

\begin{figure}
	\includegraphics[width=0.45\textwidth]{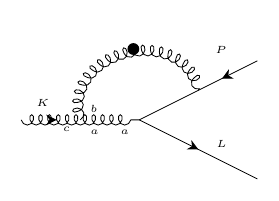}
	\includegraphics[width=0.45\textwidth]{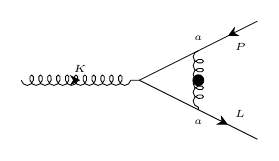}
	\caption{\label{quark_loop}Resummed quark-gluon vertex with two possible insertions of soft gluon propagators with explicit color indices.}
\end{figure}
We have all the ingredients to tackle the gluon self-energy from quark loop. The spin configuration is the same as the photon self-energy, and the interaction is similar to the pure glue case, except with different color factors. In the left diagram of Fig.~\ref{quark_loop}, the quark-gluon vertex contributes $2i P^\m$ and the 3-gluon vertex contributes $-2K^\n$. The color algebra reads
\begin{align}
t^at^bf_{abc}=\frac{i}{2}C_At^c.
\end{align}
Combining these factors, we have
\begin{align}
2iP^\m(-2K^\n)\frac{i}{2}C_A=\frac{C_A}{2}4P^\m K^\n\simeq \frac{C_A}{2}4\ppr\kpr\he^\m\he^\n.
\end{align}
This is exactly what we would have in a pure glue case with momentum exchange between $P$ and $K$.
For the right diagram, two quark gluon vertex contribute $2iP^\m$ and $2iL^\n$. The color algebra reads
\begin{align}
t^at^ct^a=\(C_F-\frac{1}{2}C_A\)t^c.
\end{align}
Similarly, the combined factor is $\(C_F-\frac{1}{2}C_A\)(-4\ppr\lpr\he^\m\he^\n)$. This is also the same as the pure glue case except with the replacement in the color factor $\frac{C_A}{2}\to \(C_F-\frac{1}{2}C_A\)$.
We can then write down the integral for the resummed vertex
\begin{align}
&\(i\d E+2\frac{\G_q}{2}+\frac{\G_g}{2}\)F(P,-L,Q)=I_0(P,-L,K)+g^2\int_Q 2\pi\d(q_0-q_\pr) D_{rr,\m\n}^*(Q)\he^\m\he^\n\Big[\no
&\(C_F-\frac{1}{2}C_A\)F(P+Q,-L-Q,K)+\frac{C_A}{2}F(P+Q,-L,K-Q)+\frac{C_A}{2}F(P,-L-Q,K+Q)\Big].
\end{align}
The corresponding 3-body pairwise interaction is asymmetric with the weight characterized by different color factors.
Using explicit representations \eqref{Gamma_q} and \eqref{Gamma_g}, we can rewrite the above as
\begin{align}\label{F_qg}
&i\d E F(P,-L,Q)=I_0(P,-L,K)+g^2\int_Q 2\pi\d(q_0-q_\pr) D_{rr,\m\n}^*(Q)\he^\m\he^\n\Big[\no
&\(C_F-\frac{1}{2}C_A\)F(P+Q,-L-Q,K)+\frac{C_A}{2}F(P+Q,-L,K-Q)+\frac{C_A}{2}F(P,-L-Q,K+Q)\Big].
\end{align}
The spin configurations of $F$ and $I_0$ are implicit in the above. Since the spin configurations are orthogonal, we can simplify \eqref{F_qg} further for three spin configurations separately. Let's still use the $\Gamma_s^{A,B,C}$ as defined in \eqref{Gamma_ABC}. Note that there is still one independent momentum, which is taken to be $\bh$, thus we expect the functions $\G_s^{A,B,C}$ to be a function of $\bh$. For $\G_s^A$, we have
\begin{align}
&i\d E h^s\G_s^A(\bh)=h^s+g^2\int_Q 2\pi\d(q_0-q_\pr) D_{rr,\m\n}^*(Q)\he^\m\he^\n\Big[\(C_F-\frac{1}{2}C_A\)(h+\kpr { q}_\pp)^s\G_s^A(\bh+\kpr {\bf q}_\pp)\no
&+\frac{C_A}{2}(h+\ppr { q}_\pp)^s\G_s^A(\bh+\ppr {\bf q}_\pp)+\frac{C_A}{2}(h+\lpr { q}_\pp)^s\G_s^A(\bh+\lpr {\bf q}_\pp)-\(C_F+\frac{1}{2}C_A\)h^s\G_s^A(\bh)\Big].
\end{align}
$\G_s^B$ satisfies the same equation with the replacement $s\to-s$ in the superscripts. The subscript $s$ can be dropped because $D_{rr,\m\n}^*(Q)\he^\m\he^\n$ is unpolarized. The two degenerate equations are equivalent to \cite{Arnold:2008iy} with the identification $\bh\G^A=\bh\G^B=\frac{1}{2}{\bf F}(h)^{\text{AMY}}$.
For $\G_s^C$, we have
\begin{align}
&i\d E\zeta_s(\bh)=1+g^2\int_Q 2\p\d(q_0-q_\pr)D_{rr,\m\n}^*(Q)\he^\m\he^\n\Big[\(C_F-\frac{1}{2}C_A\)\G_s^C(\bh+\kpr {\bf q}_\pp)\no
&+\frac{C_A}{2}\G_s^C(\bh+\ppr {\bf q}_\pp)+\frac{C_A}{2}\G_s^A(\bh+\lpr {\bf q}_\pp)-\(C_F+\frac{1}{2}C_A\)\G_s^C(\bh)\Big].
\end{align}

It remains to express the loss component of inelastic collision term in terms of $\G$ and $\zeta$. Following similar procedure as in Sec.~\ref{sec:3g}, we obtain
\begin{align}\label{loss_ql}
&\int_{k_0}\P^{\m\n>}(K)D_{\n\m}^<(K)=(-1)g^2C_F\int_{\ppr,\ppp} f(\ppr)f(\kpr)f(-\lpr)\frac{1}{-8\ppr\lpr\kpr}2\text{Re}[F(P,-L,K)I_0(P,-L,K)^*]\no
&=-g^2C_F\int_{\ppr,h} f(\ppr)f(\kpr)(1+f(\lpr))\frac{1}{\ppr\lpr \kpr^3}\text{Re}[\frac{\lpr^2+\ppr^2}{\lpr\ppr}\frac{h^2}{\kpr^2}\G+\frac{m^2\kpr^2}{\lpr\ppr}\zeta],
\end{align}
where $\G=\G_A=\G_B$ and $\zeta=\zeta_s$.
The gain term is obtained by the change of the phase space distribution $f(\ppr)f(\kpr)(1+f(\lpr))\to f(\lpr)(1+f(\ppr))(1+f(\kpr))$ in the integrand.

\subsubsection{Quark inelastic collision}

\begin{figure}
	\includegraphics[width=0.45\textwidth]{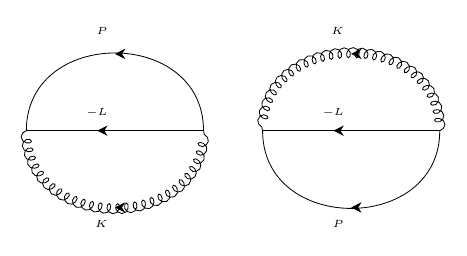}
	\caption{\label{qg_equiv}Equivalence of inelastic gluon self-energy diagram from quark loop and inelastic quark self-energy diagram.}
\end{figure}
The inelastic collision term for quark requires only a little further effort. To see that, we first note the diagrammatic analysis in Sec.~\ref{sec:diagram} is independent of spin, thus is equally applicable for the quark case. It follows that quark loop contribution to gluon inelastic collision is closed related to quark inelastic collision, as shown in Fig.~\ref{qg_equiv}. Since the diagrammatic relation holds for unintegrated collision terms, we need to undo the integrations in \eqref{loss_ql} to obtain the following explicit relation
\begin{align}
\int_{k_0}\times\!\!\!\!\!\!\int_{\ppr,\ppp}\P^{\m\n>}(K)D_{\n\m}^<(K)=
\int_{p_0}\times\!\!\!\!\!\!\int_{\kpr,\kpp}\tr[\S^>(P)S^<(P)],
\end{align}
with $\times\!\!\!\!\!\int_{x}$ indicating not integrating over $x$.
A more useful relation is obtained by integrating over the transverse momentum, which resums multiple interactions. Since there is only one independent transverse momentum $h=\kpr\ppp-\ppr\kpp$, we can convert between the measures $d^2\ppp$ and $d^2\kpp$ as $d^2\ppp=\frac{d^2h}{\kpr^2}=\frac{\ppr^2}{\kpr^2}d^2\kpp$. It follows then
\begin{align}
\int_{p_0}\times\!\!\!\!\!\!\int_{\kpr}\tr[\S^>(P)S^<(P)]=\frac{\kpr^2}{\ppr^2}\int_{k_0}\times\!\!\!\!\!\!\int_{\ppr}\P^{\m\n>}(K)D_{\n\m}^<(K).
\end{align}
We already know $\int_{k_0}\times\!\!\!\!\!\int_{\ppr}\P^{\m\n>}(K)D_{\n\m}^<(K)$ from \eqref{loss_ql}. We then immediately have
\begin{align}
\int_{p_0}\tr[\S^>(P)S^<(P)]=-g^2C_F\int_{\kpr,h} f(\ppr)f(\kpr)(1+f(\lpr))\frac{1}{\ppr^3\lpr \kpr}\text{Re}[\frac{\lpr^2+\ppr^2}{\lpr\ppr}\frac{h^2}{\kpr^2}\G+\frac{m^2\kpr^2}{\lpr\ppr}\zeta].
\end{align}

\section{Quantum kinetic equation and spin polarization}\label{qke_pol}

With $S^{<(0)}$ and $D^{<(0)}$ determined from spin averaged kinetic equations, we are ready to consider $S^{<(1)}$ and $D^{<(1)}$, which encodes spin polarization of quarks and gluons respectively. This requires expanding the KB equations to $O(\pd)$. This has been done systematically in \cite{Yang:2020hri} and \cite{Hattori:2020gqh} for generic self-energies. We shall proceed a slightly different way: since we specialize to QCD interaction and aim at application to spin polarization phenomena in QGP, we will perform an analysis by combining a power counting of self-energies in coupling and a classification of hydrodynamic gradients. This will lead to significant simplified contributions to $S^{<(1)}$ and $D^{<(1)}$. We start by expanding \eqref{KB_q} as
\begin{align}\label{KB_q_pd}
&\frac{i}{2}\slashed{\pd}S^{<(0)}+\(\slashed{P}-m\)S^{<(1)}=\frac{i}{2}\(\S^{>(0)}S^{<(0)}-\S^{<(0)}S^{>(0)}\)-\frac{1}{4}\(\{\S^{>(0)},S^{<(0)}\}_\pb-\{\S^{<(0)},S^{>(0)}\}_\pb\),\no
&\frac{i}{2}\slashed{\pd}S^{<(1)}+\(\slashed{P}-m\)S^{<(2)}=\frac{i}{2}\(\S^{>(1)}S^{<(0)}-\S^{<(1)}S^{>(0)}+\S^{>(0)}S^{<(1)}-\S^{<(0)}S^{>(1)}\)+\text{P.B.},
\end{align}
where self-energy modification to dispersion terms have been dropped. The first equation is not formally the same order in gradient. In particular the trace (diagonal components) of the first terms on both sides giving the Boltzmann equation match terms at different superficial orders in gradient. The LHS for a generic hydrodynamic gradient acting on local equilibrium distribution in $S^{<(0)}$ produces a term at $O(g^0\pd)$. A nonvanishing collision term on the RHS is parametrically $O(g^4\d f_q)$, with $\d f_q$ being the deviation of local distribution from equilibrium. Matching the two sides, we obtain $\d f_q\sim O(\pd/g^4)$. Physically, the presence of hydrodynamic sources drives the local equilibrium distribution into a steady state distribution\footnote{We don't have to use the deviation on the LHS as it is higher order in gradient}. It is in this steady state that the corresponding transport coefficient is evaluated, which is parametrically $O(1/g^4)$ \cite{Arnold:2000dr,Arnold:2003zc}. An exception to the generic case is the hydrodynamic gradient of vorticity, for which the gradient on the LHS vanishes identically and the system remains in local equilibrium distribution with vanishing collision term.

The off-diagonal components of the first equation in \eqref{KB_q_pd} constrain $S^{<(1)}$. Ignoring the Poisson bracket terms, which are higher order in $\pd$, we obtain \cite{Lin:2021mvw}
\begin{align}\label{S1}
S^{<(1)}(P)=\g^5\g_\m{\cal A}^\m+\frac{i[\g_\m,\g_\n]}{4}{\cal S}^{\m\n},
\end{align}
where
\begin{align}\label{AS}
&{\cal A}^\m=-2\p\e(P\cdot u)\frac{\e^{\m\n\r\s}P_\r u_\s{\cal D}_\n f_q}{2(P\cdot u+m)}\d(P^2-m^2),\no
&{\cal S}^{\m\n}=-2\p\e(P\cdot u)\frac{{\cal D}_{[\m} P_{\n]}f_q-m u_{[\m} {\cal D}_{\n]}f_q-P_{[\m} u_{\n]}{\cal D}_m}{2(P\cdot u+m)}\d(P^2-m^2).
\end{align}
The collisional effect is absorbed in the definitions ${\cal D}_\n=\pd_\n-\S_\n^>-\S_\n^<\frac{1-f_q}{f_q}$ and ${\cal D}_m=\S_m^>+\S_m^<\frac{1-f_q}{f_q}$ with $\S_\n^{>/<}=\frac{1}{4}\tr[\S^{>/<}\g_\n]$ and $\S_m^{>/<}=\frac{1}{4}\tr[\S^{>/<}]$.
Crucially the collision contribution in ${\cal D}_\nu$ is parametrically the same as $\pd_\n$ because $-\S_\n^>-\S_\n^<\frac{1-f_q}{f_q}\sim g^4\d f_q\sim\pd$. This contribution has been calculated explicitly in \cite{Lin:2022tma,Lin:2024zik,Wang:2024lis,Fang:2023bbw,Fang:2024vds} for shear source. Obviously such contribution doesn't exist for vorticity as the system remains in local equilibrium.
The solution \eqref{S1} is not unique, but subject to addition of homogeneous solutions. In principle the full solution can be fixed by the second equation of \eqref{KB_q_pd}, which is a more complicated dynamical equation for $S^{<(1)}$. An explicit determination of the homogeneous solution has been worked out for shear source in \cite{Wang:2024lis}. However, the situation simplifies significantly when we consider the massless limit. In this case, the spin of quark is enslaved by its momentum, with the homogeneous solution corresponding to the difference in the distribution of two helicities. We shall discuss the cases of non-vortical and vortical gradients separately. The response of two helicities to non-vortical sources are the same, since spin couples to vorticity only. The homogeneous solution is expected to vanish, thus \eqref{S1} is the complete solution. For vortical source, the difference in the distribution of the two helicities is given by spin-vorticity coupling in the free theory limit. The collisional contribution in \eqref{S1} for this case simply vanishes in local equilibrium. Other non-collisional interaction contribution is possible but are suppressed by additional powers of coupling \cite{Lin:2024svh,Fang:2025pzy}. Indeed, the features reasoned above is supported by analysis using frame independence in chiral kinetic theory in \cite{Wang:2024lis}. It is possible to extend the applicability of \eqref{S1} to small mass regime: the analysis in \cite{Wang:2024lis} suggests the mass effect is parametrically $O(m\pd)$. With our parametric choice $m\sim gT$, the mass effect may be ignored. Let's summarize the approximate expression for axial component covering both vortical and non-vortical sources
\begin{align}\label{S1_all}
S^{<(1)}=-2\p\e(P\cdot u)\big[P^\m f_q^A(P)+\frac{\e^{\m\n\r\s}P_\r u_\s{\cal D}_\n f_q}{2P\cdot u}\big]\d(P^2-m^2),
\end{align}
with $f_q^A=\frac{P\cdot\O}{2P\cdot u}\frac{\pd f_q}{\pd(P\cdot u)}$ and $\O^\m=\frac{1}{2}\e^{\m\n\r\s}u_\n\pd_\r u_\s$ is the vorticity vector.

The gluon case is analyzed similarly. Dropping the self-energy modification to dispersion and the Poisson bracket terms, we have the following equation expanded in gradient
\begin{align}\label{KB_g_pd}
&\(-P^2\h^{\m\n}+P^\m P^\n-\frac{1}{\x}P^{\m\a}P^{\n\b}P_\a P_\b\)D^{<(1)}_{\n\r}+\frac{i}{2}\(-2P\cdot\pd \h^{\m\n}+(\pd^\m P^\n+\pd^\n P^\m)
-\frac{1}{\x}P^{\m\a}P^{\n\b}(\pd_\a P_\b+\pd_\b P_\a)\)D^{<(0)}_{\n\r}\no
&=\frac{i}{2}\(\P^{\m\n>(0)}D_{\n\r}^{<(0)}-\P^{\m\n<(0)}D_{\n\r}^{>(0)}\).
\end{align}
Note that we have used the transverse component of this equation to obtain the Boltzmann equation for gluons. The remaining components serve as a constraint for $D_{\n\r}^{<(1)}$. The solution has been obtained in \cite{Lin:2021mvw} as
\begin{align}\label{D1}
&D_{\l\r}^{<(1)}(P)=-2\p\e(P\cdot u)\d(P^2)\frac{iP_{\l\a}P^{\n\b}P^\a\pd_\b P_{\n\r}^T f_g(P)}{2(-P^2+(P\cdot u)^2)}+2\p\e(P\cdot u)\d(P^2)P_{\n\r}^T\times\no
&\frac{i u_\l u_\m\(\P^{\m\n>(0)}f_g(P)-\P^{\m\n<(0)}(1+f_g(P))\)}{2(-P^2+(P\cdot u)^2)}-(\l\leftrightarrow\r),
\end{align}
with $\x$ set to zero for strict Coulomb gauge.
It is not difficult to see that \eqref{D1} has one transverse index only. The first term due to gradient corresponds to mixing between transverse polarization to longitudinal polarization, and the second term due to collisional contribution corresponds to mixing transverse polarization and time-like polarization. As is in the case of quark, the collisional term proportional to $\P^{\m\n>(0)}f_g-\P^{\m\n<(0)}(1+f_g)$ is parametrically $O(g^0\pd)$ and vanishes in local equilibrium. With the same logic as the quark case, we discuss separately vortical and non-vortical hydrodynamic sources. For the former, there is an additional homogeneous solution given by spin-vorticity coupling \cite{Huang:2020kik,Hattori:2020gqh} with no collisional contribution. For the latter, the collisional contribution represents a steady state effect but there is no homogeneous solution as the gluon is strictly massless. The two cases can be unified into the following expression
\begin{align}\label{D1_all}
&D_{\l\r}^{<(1)}(P)=-2\p\e(P\cdot u)\d(P^2)\frac{iP_{\l\a}P^{\n\b}P^\a\pd_\b P_{\n\r}^T f_g(P)}{2(-P^2+(P\cdot u)^2)}+2\p\e(P\cdot u)\d(P^2)P_{\n\r}^T\times\no
&\frac{i u_\l u_\m\(\P^{\m\n>(0)}f_g(P)-\P^{\m\n<(0)}(1+f_g(P))\)}{2(-P^2+(P\cdot u)^2)}-(\l\leftrightarrow\r)-2\p \e(P\cdot u)\frac{i\e_{\l\r\a\b}P^\a u^\b}{P\cdot u}f_A(P)\d(P^2),
\end{align}
with $f_A=\frac{P\cdot\O}{P\cdot u}\frac{\pd f_g}{\pd(P\cdot u)}$.

\section{Gauge dependencies}\label{gauge_dep}

We have derived the quantum kinetic equations in Coulomb gauge. It is natural to discuss the gauge dependencies. In the equilibrium limit, quarks and transverse gluons are the only DOF, with gauge independent thermal distributions. Out of equilibrium, the distribution functions satisfy the Boltzmann equation. If the collision kernel is gauge independent, the distribution functions are also gauge independent. We now argue this is indeed the case with the approximation adopted: the elementary square amplitudes for both elastic and inelastic collisions are expected to be gauge independent, as the initial and final states are physical. However, the situation becomes subtle in medium, where effects of screening and modification to dispersions come into play. These effects arising from one-loop (or effective one-loop) self-energies are in general gauge dependent. But with self-energies in the HTL approximation adopted in this work, these effects are still gauge independent. It follows that the distribution functions are also gauge independent.

The two-loop and multiple-loop self-energies giving rise to collision term have hard external momenta are obviously beyond HTL. It is remarkable that when diagonal components of collision term extracted by trace/transverse projection for quark/gluon leads to gauge independent collision term. The non-diagonal components entering \eqref{S1_all} and \eqref{D1_all} are in general gauge dependent, see explicit example of fermion spin polarization in \cite{Lin:2022tma}. The gauge dependence is expected to cancel in final hadron polarization. This implies a careful choice of hadron structure model is required in order to achieve a gauge independent hadron polarization. Naive phenomenological model is inadequate for this purpose.

\section{Summary and Outlook}\label{sum_out}

We have developed a quantum kinetic theory for QCD based on gradient expanded Kadanoff-Baym equation. At lowest order in gradient, it reduces to the spin averaged Boltzmann equation with both elastic and inelastic collisions. At next order in gradient, the quarks and gluons gain spin polarization in response to the gradient, with collisional contribution contained in the off-diagonal component of the lowest order collision term. The spin polarization behaves differently in vortical and non-vortical gradients: to leading order in the coupling, the vorticity induced polarization is saturated by the free theory result, while for non-vortical gradient, the collisional contribution exists in general and is parametrically the same order as the free theory counterpart.

We also discuss the gauge dependencies of the quantum kinetic theory: at lowest order, the equation is gauge independent in the HTL approximation we use. However, at next order in gradient, gauge dependence exists in general. In particular, it shows up in collisional contribution to spin polarization, suggesting care choice of hadron structure model is required. Despite of the caveat, it is clearly desirable to apply the present framework to study more realistic collisional contribution to spin polarization in QGP. 

Another notable observation is the description of inelastic collision in spin basis. It shows change of orbital angular momentum and spin in the collision while conserves the total, offering a mechanism for spin-orbital conversion. While this mechanism doesn't lead to a physical consequence as our assumption that distrbutions are unpolarized at lowest order in gradient. This assumption can certainly be lifted. It would be interesting to explore the case of polarized distribution, in which the mechanism is expected to play a role.

\section*{Acknowledgements}

I thank Defu Hou and Qun Wang for useful discussions. I also thank Ziyue Wang for collaboration on related works. This work is in part supported by NSFC under Grant Nos 12475148 and 12075328.

\appendix

\section{Contour integration}

We shall perform the following integral
\begin{align}
\int\frac{p_0}{2\p}\frac{k_0}{2\p}D_{ra}(P)D_{ra}(-L)D_{ra}(K)
\end{align}
The explicit form of the retarded propagator is given by
\begin{align}\label{retarded}
D_{ra}(P)=\frac{i}{(p_0+i\G/2)^2-E_p^2},
\end{align}
with $E_p=\ppr+\frac{p_\pp^2+m_g^2}{2\ppr}$
and similarly for the other two. We first perform the $p_0$-integration. With $l_0=p_0+k_0$, we easily obtain
\begin{align}\label{int_PL}
\int_{p_0}D_{ra}(P)D_{ra}(-L)\simeq\frac{i}{4\ppr\lpr(\d E'+k_0-\kpr-i\G)},
\end{align}
with $\d E'=\frac{p_\pp^2+m_g^2}{2\ppr}+\frac{l_\pp^2+m_g^2}{-2\lpr}$. The remaining $k_0$-integration is done by keeping the pole of $D_{ra}(K)$, which gives
\begin{align}
\int_{p_0,k_0}D_{ra}(P)D_{ra}(-L)D_{ra}(K)=\frac{i}{8\ppr\lpr\kpr(\d E-3i\G/2)},
\end{align}
with $\d E=\frac{p_\pp^2+m_g^2}{2\ppr}+\frac{l_\pp^2+m_g^2}{-2\lpr}+\frac{k_\pp^2+m_g^2}{2\kpr}$.
We also need a three-fold integral in the term corresponding to soft gluon exchange, which we perform as follows
\begin{align}
&\int_{p_0,k_0,q_0}D_{ra}(P+Q)D_{ra}(-L-Q)D_{ra}(K)D_{ra}(P)D_{ra}(-L)\no
&=\int_{p_0+q_0,k_0}D_{ra}(P+Q)D_{ra}(-L-Q)D_{ra}(K)\int_{p_0}D_{ra}(P)D_{ra}(-L)\no
&\simeq\frac{1}{-8\ppr\lpr\kpr(i\d E+3\G/2)}\int_{p_0}D_{ra}(P)D_{ra}(-L).
\end{align}
In the first level of integration, we have fixed the poles at $p_0+q_0=\ppr+q_\pr+\frac{(p_\pp+q_\pp)^2+m_g^2}{2\ppr}-i\frac{\G}{2}$ and $k_0=\kpr+\frac{\k_pp^2+m_g^2}{2\kpr}-i\frac{\G}{2}$. With $q_0$ undertermined, we still have $p_0$ independent from $p_0+q_0$, but $l_0=p_0+\kpr++\frac{\k_pp^2+m_g^2}{2\kpr}-i\frac{\G}{2}$. This allows us to perform the remaining integration as
\begin{align}
&\int_{p_0}D_{ra}(P)D_{ra}(-L)\simeq\frac{1}{4\ppr\lpr}\int_{p_0}\frac{i}{p_0+\frac{i}{2}\G-E_p}\frac{i}{l_0-\frac{i}{2}\G-E_l}\no
&\simeq\frac{-1}{4\ppr\lpr}\frac{1}{i\d E+3\G/2},
\end{align}
using the value of $l_0$ from the pole.

\bibliographystyle{unsrt}
\bibliography{QCD_QKT.bib}

\end{document}